\documentclass[preprint,aps,showpacs]{revtex4}

\usepackage{graphicx}
\usepackage{dcolumn}
\usepackage{bm}
\usepackage{epsfig}
\usepackage{psfrag}

\usepackage{dsfont}
\usepackage{cancel}
\usepackage{amsmath}
\usepackage{amssymb}
\usepackage{tabularx}
\usepackage{color}
\usepackage{fontenc}
\usepackage{times}
\usepackage{subfigure}
\usepackage{fancybox}

\def\gev{\,{\rm GeV}}

\newcommand{\nn}{\nonumber}
\newcommand{\be}{\begin{equation}}
\newcommand{\ee}{\end{equation}}
\newcommand{\best}{\begin{equation*}}
\newcommand{\eest}{\end{equation*}}
\newcommand{\ba}{\begin{eqnarray}}
\newcommand{\ea}{\end{eqnarray}}
\newcommand{\bea}{\begin{eqnarray}}
\newcommand{\eea}{\end{eqnarray}}

\def\={\,=\,}
\def\:={\,:=\,}

\newcommand{\req}[1]{(\ref{#1})}

\newcommand{\tr}[1]{ {\bf #1}_\perp}

\def\vd{{\bm \Delta}_\perp}
\def\vdd{ {\bm \Delta}_\perp^2}
\newcommand{\vk}{{\bm k}_{\perp}}

\newcommand{\sla}{\hspace*{-0.18cm}/}

\def\ci{\cite}

\def\xb{\bar{x}}

\def\ourprocess{p\bar{p} \, \to \, \overline{D^0} D^0}

\newcommand{\VEC}[1]{\mathbf{#1}}

\begin{document}

\title{Double handbag description of proton-antiproton annihilation\\ into a heavy meson pair}
\author{Alexander T.\ Goritschnig}
\email{alexander.goritschnig@uni-graz.at}
\author{Bernard Pire}
\email{bernard.pire@cpht.polytechnique.fr} \affiliation{Centre de Physique Th\' eorique, 
\' Ecole Polytechnique, CNRS, 91128 Palaiseau, France}
\author{Wolfgang Schweiger}%
\email{wolfgang.schweiger@uni-graz.at} \affiliation{ Institut
f\"ur Physik, Universit\"at Graz, A-8010 Graz, Austria}
\date{\today}


\begin{abstract}
We propose to describe the process $ p \bar{p} \,\to\, \overline{D^0} D^0$ in a perturbative QCD motivated framework where a double-handbag hard process $ u d \bar u \bar d \to \bar{c} c$ factorizes from transition distribution amplitudes, which are quasiforward hadronic matrix elements of $\Psi_q \Psi_q \Psi_c $ operators, where q denotes light quarks and c denotes the heavy quark. We advocate that the charm-quark mass acts as the large scale allowing this factorization. We calculate this process in the simplified framework of the scalar diquark model and present the expected cross sections for the PANDA experiment at GSI-FAIR.
\end{abstract}


\maketitle

\section{Introduction}
\label{sec:intro}

The collinear factorization framework allows us to calculate a number of hard exclusive amplitudes
in terms of perturbatively calculable coefficient functions and nonperturbative hadronic
matrix elements of light-cone operators. The prime example is the calculation of
the deeply virtual Compton-scattering amplitude in the handbag approximation
with generalized parton distributions,
nonforward matrix elements of a quark-antiquark nonlocal operator
$\Psi(z)\overline{\Psi}(0)$ between an incoming and an outgoing baryon state.
Strictly speaking, this description is only valid in a restricted kinematical region,
called the generalized Bjorken scaling region, for a few specific reactions,
and in the leading-twist approximation. It is, however, suggestive to extend
this framework to the description of other reactions
where the presence of a hard scale seems to justify the factorization of a short-distance dominated
partonic subprocess from long-distance hadronic matrix elements. Such an extension has,
in particular, been proposed in Ref.~\cite{Goritschnig:2009sq} for the reaction
$p\bar{p} \,\to\, \Lambda_c\bar{\Lambda}_c$ with nucleon to charmed baryon generalized parton distributions.
The extension of the collinear factorization framework to the backward region of deeply virtual 
Compton-scattering and deep exclusive meson production
\cite{Pire:2004ie, Lansberg:2012ha}
leads to the definition of transition distribution amplitudes (TDAs) as nonforward matrix elements
of a three-quark nonlocal operator $\Psi(z)\Psi(y)\Psi(0)$ between an incoming and an outgoing
state carrying a different baryon number. Here, too, this description is likely to be valid in a restricted
kinematical region, for a few specific reactions, and in the leading twist approximation.
We propose here to extend the approach of Ref.~\cite{Goritschnig:2009sq} to the reaction
$p\bar{p} \,\to\, \overline{D^0} D^0$ which will be measured with the
PANDA \cite{PANDA} detector at GSI-FAIR.
For this process the baryon number exchanged in the t-channel implies
that hadronic matrix elements with $\Psi(z)\Psi(y)\Psi(0)$ operators enter the game.
Let us stress that we have no proof of the validity of this approximation but
take it as an assumption to be confronted with experimental data.
For this approach to be testable, one needs to model the occurring
nucleon to charmed meson TDAs.
In contrast to the $N\to \pi$ TDAs, which have been much discussed \cite{PSS}, we do not have any soft meson limit to normalize these TDAs. We will rather use an overlap representation in the spirit of Ref.~\cite{Diehl:2000xz}.

\section{Hadron Kinematics}
\label{sec:kinematics}

%
\begin{figure}[h]
\begin{center}
\includegraphics[width=.47\textwidth, clip=true]{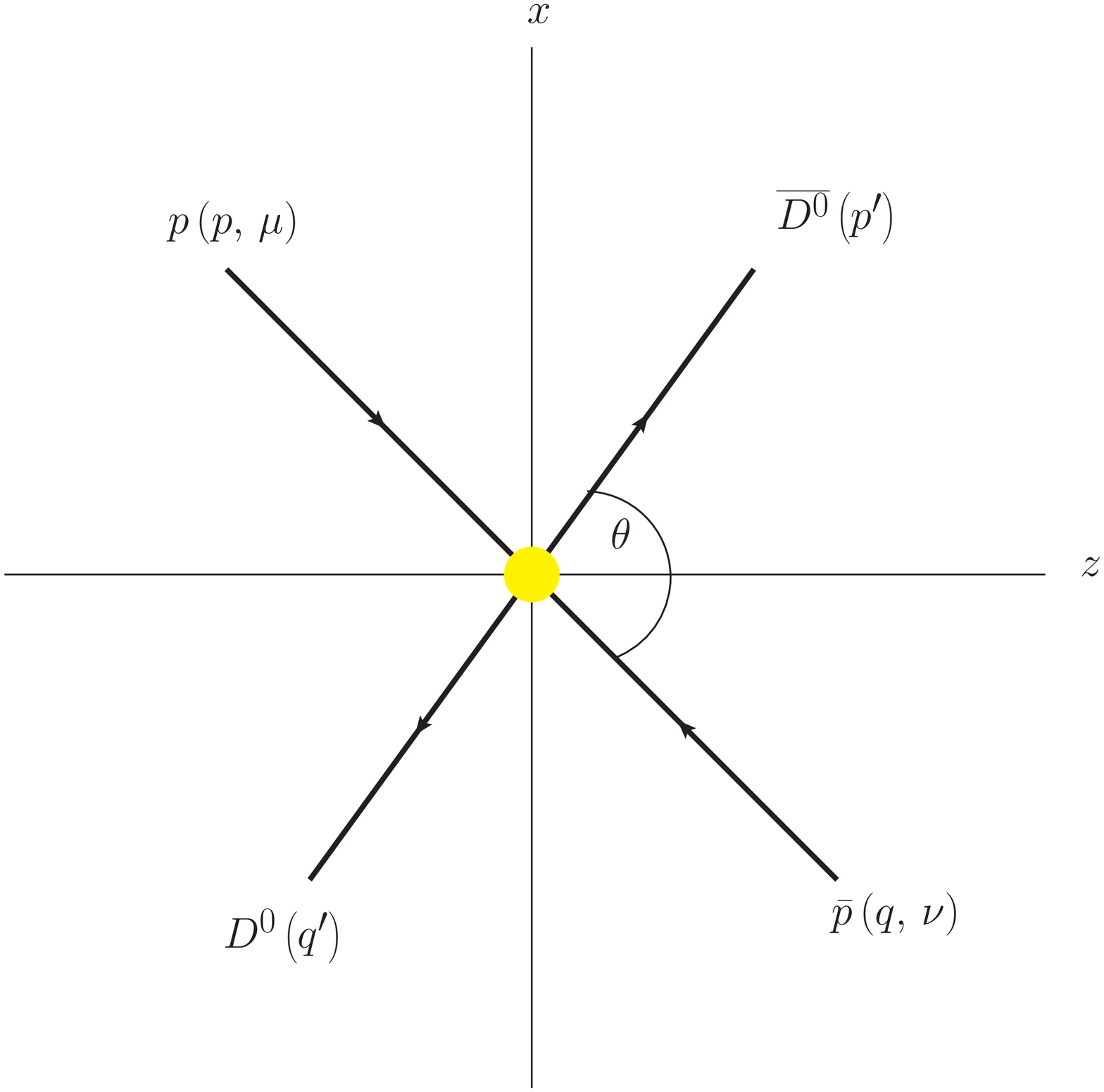}
\end{center}
\caption{Kinematics of $\ourprocess$ in the symmetric CMS.}
\label{fig:kinemtics}
\end{figure}

The kinematical situation for $\ourprocess$ scattering
is sketched in Fig.~\ref{fig:kinemtics}.
The momenta and helicities of the incoming proton and antiproton are
denoted by $p$, $\mu$ and $q$, $\nu$ and the momenta of the outgoing
$\overline{D^0}$ and $D^0$  by $p'$ and $q'$, respectively. \
The mass of the proton is denoted by $m$ and that of the $D^0$ by $M$.
We choose a symmetric center-of-momentum system (CMS) in which
the longitudinal direction is defined by the average momentum of \
the incoming proton and the outgoing $\overline{D^0}$, respectively.
The transverse momentum transfer is symmetrically shared
between the incoming and outgoing hadrons.
In light-cone coordinates the hadronic momenta are parameterized as follows,
\be
\begin{split}
p &=\left[(1+\xi)\bar{p}^{\,+},\, \frac{m^2+\vdd/4}{2(1+\xi)\bar{p}^{\,+}},\,
 -\frac{\vd}2 \right ]\,, \quad
p^\prime=\left[(1-\xi )\bar{p}^{\,+},\, \frac{M^2+\vdd/4}{2(1-\xi)\bar{p}^{\,+}},
                             \,+\frac{\vd}2\right ]\,, \\
q &=\left[\frac{m^2+\vdd/4}{2(1+\xi)\bar{p}^{\,+}},\,(1+\xi)\bar{p}^{\,+},\,
 +\frac{\vd}2\right]\,, \quad
q^\prime=\left[\frac{M^2+\vdd/4}{2(1-\xi)\bar{p}^{\,+}},\,(1-\xi )\bar{p}^{\,+},\,
-\frac{\vd}2 \right]\,,\\
\label{def-momenta-ji}
\end{split}
\ee
where we have introduced sums and differences of the hadron momenta,
\be
\bar{p} \ := \ \frac12(p+p^\prime)\,, \qquad \bar{q} \ := \ \frac12(q+q^\prime)\,
\qquad \text{and}\qquad
\Delta \ := \ p^\prime-p=q-q^\prime\,.
\label{sum-and-diff}
\ee
The minus momentum components can be obtained by using the on-mass shell conditions
$p^2 = q^2 = m^2$ and $p^{\prime 2} = q^{\prime 2} = M^2$.
The skewness parameter $\xi$ gives the relative momentum transfer in the plus direction, i.e.,
\be
\xi \ := \ \frac{p^{+}-p^{\prime +}}{p^{+}+p^{\prime+}}
      = - \frac{\Delta^+}{2\bar{p}^{\,+}} \,.
\label{skewness}
\ee
The Mandelstam variable $s$ is given by
\be
s \=\ \left( p + q \right)^2
  \=\ \left( p^\prime + q^\prime \right)^2 \,.
\ee
In order to produce a $\overline{D^0}\,D^0$ pair,
$s$ must be larger than $4 M^2$.
The remaining Mandelstam variables, $t$ and $u$, read
\be
t \=\ \Delta^2
  \=\ \left( p^\prime - p \right)^2
  \=\ \left( q - q^\prime \right)^2
\ee
and
\be
u \=\ \left( q^\prime - p \right)^2
  \=\ \left( p^\prime - q \right)^2 \,,
\ee
so that $s+t+u = 2 M^2 + 2 m^2$.
For later convenience we also introduce the abbreviations
\be
\Lambda_m \:=\ \sqrt{1-4m^2/s}
\qquad\text{and}\qquad
\Lambda_M \:=\ \sqrt{1-4M^2/s} \,.
\ee
For further relations between the kinematical quantities, see Appendix~\ref{app-kinematics}.
%

\section{Double Handbag Mechanism}
\label{sec:DHM}

\begin{figure}[h]
\begin{center}
\includegraphics[width=0.70\textwidth, clip=true]{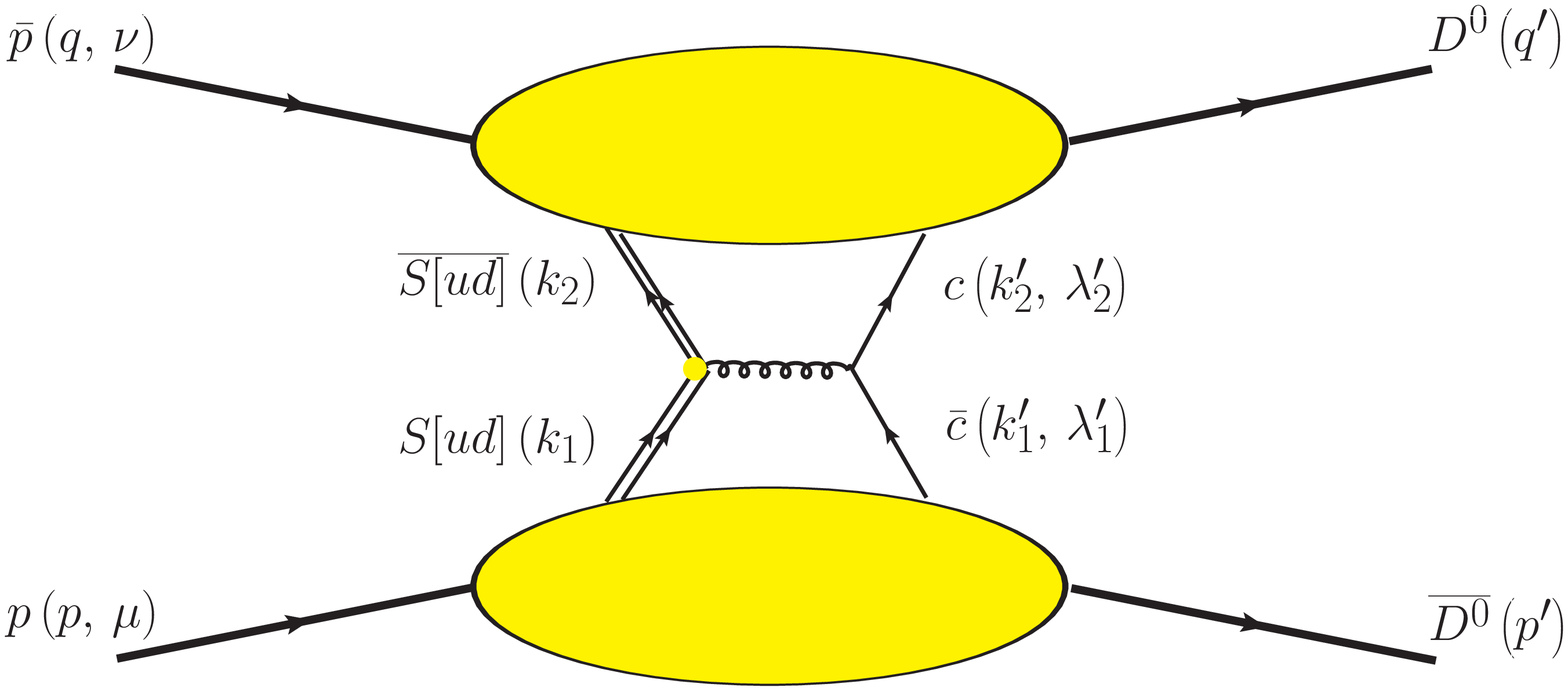}
\end{center}
\caption{The handbag contribution to the process $\ourprocess$.
  The momenta and helicities of the baryons and quarks are specified.}
\label{fig:handbag}
\end{figure}

The double handbag mechanism which we use to describe
$\ourprocess$
is shown in  Fig.\ \ref{fig:handbag}.
It is understood that the proton emits an $S[ud]$ diquark with momentum $k_1$
and the antiproton a $\overline{S[ud]}$-diquark with momentum $k_2$.
They undergo a scattering with each other, i.e. they annihilate in our case
into a gluon which subsequently decays into the heavy $\bar{c}c$ pair.
Those produced heavy partons, characterized by $k^\prime_1$, $\lambda^\prime_1$
and $k^\prime_2$, $\lambda^\prime_2$, are reabsorbed by the remnants
of the proton and the antiproton to form the $\overline{D^0}$ and the $D^0$, respectively. 
One could, of course, also think of vector-diquark configurations in the proton
and $V\left[ud\right]\overline{V\left[ud\right]}$ annihilation to produce the $\bar{c}c$ pair.
But in common diquark models of the proton it is usually assumed that
the probability to find a $V\left[ud\right]$ diquark is smaller than the one for the $S\left[ud\right]$ diquark.
Further suppression of $V\left[ud\right]$ diquarks as compared to
$S\left[ud\right]$ diquarks occurs in hard processes via
diquark form factors at the diquark-gluon vertices \cite{Jakob:1993th}.
We thus expect that our final estimate of the $\overline{D^0}D^0$
cross section will not be drastically altered by the inclusion of
vector-diquark contributions and we stick to the simpler scalar
diquark model.
The whole hadronic four-momentum transfer $\Delta$ is also exchanged
between the active partons in the partonic subprocess
\begin{equation}
S[ud](k_1) \, \overline{S[ud]}(k_2) \,\to\,
\bar{c}(k_1^\prime , \lambda_1^\prime) \,c(k_2^\prime , \lambda_2^\prime) \,.
\label{subprocess}
\end{equation}
In Eq.~(\ref{subprocess}) we neglect the mass of the $S[ud]$ (anti)diquark,
but take into account the heavy \hbox{(anti-)} charm-quark mass $m_c$.
In order to produce the heavy $\bar{c}c$ pair,
the Mandelstam variable $\hat{s}$ of the partonic subprocess has to be
\be
\hat{s} \,\geq\, 4 m_c^2 \,,
\label{partonic-threshold}
\ee
where $4 m_c^2 \approx 6.5\,\text{GeV}^2$.
We have taken the (central) value for the charm-quark mass from the Particle Data Group \ci{PDG}, which gives $m_c\=1.275 \, \pm \, 0.025\gev$.
Thus, the heavy-quark mass $m_c$ is a natural intrinsic hard scale
which demands that the intermediate gluon has to be highly virtual.
This allows us to treat the partonic subprocess perturbatively, even at small $-t$,
by evaluating the corresponding Feynman diagram.
All the other non-active partons inside the parent hadrons
are unaffected by the hard scattering and thus act as spectators.
For the double handbag mechanism the hadronic $\ourprocess$
amplitude can be written as
\be
\begin{split}
M_{\mu \nu} = & \sum_{\rm a_i^{(\prime)}}
                \sum_{\rm \alpha_i^{\prime}}
                    \int d^4 \bar{k}_1 \theta(\bar{k}_1^{+})
                    \int \frac{d^4 z_1}{(2\pi)^4} e^{i \bar{k}_1 z_1}
                    \int d^4 \bar{k}_2 \theta(\bar{k}_2^{-})
                    \int \frac{d^4 z_2}{(2\pi)^4} e^{i \bar{k}_2 z_2}
 \\
 & \times \langle \overline{D^0} : p^\prime | \mathcal{T}\,
                  \Psi^c_{a^\prime_1\alpha^\prime_1}(-z_1/2)
                  \Phi^{S[ud]}_{a_1}(+z_1/2)
          |p: p , \mu\rangle\;
             \Tilde{H}_{a_i^{(\prime)} \alpha^{\prime}_i}(\bar{k}_1, \bar{k}_2)
             \\
 & \times \langle D^0: q^\prime | \mathcal{T}\,
                    \Phi^{S[ud]\dagger}_{a_2}(+z_2/2)
                    \bar{\Psi}^c_{a^\prime_2\alpha^\prime_2}(-z_2/2)
         |\bar{p}: q , \nu\rangle \,, \\
\label{ampl}
\end{split}
\ee
%
where the assignment of momenta, helicities, etc., can be seen in Fig.~\ref{fig:handbag}.
$a_i^{(\prime)}$ and $\alpha_i^{\prime}$ denote color and spinor indices, respectively.
In analogy to the hadronic level we have introduced the average partonic momenta
$\bar{k}_i := \left(k_i + k_i^\prime\right)/2$, $i=1,2$, of the active partons.
We note once more that the full hadronic momentum transfer is also
transferred between the active partons, i.e.
$k_1 - k_1^\prime = p - p^\prime = k_2^\prime - k_2 = q^\prime - q$. 
The hard scattering kernel,
denoted by $\Tilde{H}_{a_i^{(\prime)} \alpha^{\prime}_i}(\bar{k}_1, \bar{k}_2)$,
describes the hard $S[ud]\overline{S[ud]} \rightarrow \bar{c}c$ subprocess.
The soft part of the $p \to \overline{D^0}$ transition is encoded in the Fourier transform of
a hadronic matrix element which is a time-ordered, bilocal product of
a quark and a diquark field operator:
\be
 \int \frac{d^4 z_1}{(2\pi)^4} e^{i \bar{k}_1 z_1}
  \langle \overline{D^0} : p^\prime | {\cal T}\,
             \Psi^c_{a^\prime_1\alpha^\prime_1}(-z_1/2)
             \Phi^{S[ud]}_{a_1}(+z_1/2)
  |p : p , \mu\rangle \,.
\label{soft-Tordered-hadronic-matrix-element}
\ee
In Eq.~(\ref{soft-Tordered-hadronic-matrix-element}) $\Phi^{S[ud]}(+z_1/2)$ takes out
an $S[ud]$ diquark from the proton state $|p : p , \mu\rangle$ at the
space-time point $z_1/2$.
The $S[ud]$ diquark then takes part in the hard partonic subproces.
The $\Psi^c(-z_1/2)$ reinserts the $\bar{c}$ quark at $-z_1/2$
into the remnant of the proton which gives the desired final hadronic
$\overline{D^0}$ state $|\overline{D^0} : p^\prime \rangle$.
At this stage the appropriate time-ordering of the quark field operators
(denoted by the symbol ${\cal T}$) has to be taken into account.
The remnant of the proton,
which does not participate in the hard partonic subprocess,
constitutes the spectator system.
For the $\bar{p} \to D^0$ transition we have the Fourier transform
\be
 \int \frac{d^4 z_2}{(2\pi)^4} e^{i \bar{k}_2 z_2}
  \langle D^0 : q^\prime | {\cal T}\,
                \Phi^{S[ud]\dagger}_{a_2}(+z_2/2)
                \overline{\Psi}^c_{a^\prime_2\alpha^\prime_2}(-z_2/2)
  |\bar{p} : q , \nu\rangle \,,
\label{soft-Tordered-anti-hadronic-matrix-element}
\ee
which can be interpreted in a way analogous to Eq.~(\ref{soft-Tordered-hadronic-matrix-element}).
The $\ourprocess$ amplitude (\ref{ampl}) is
thus a convolution of a hard scattering kernel with hadronic matrix elements
Fourier transformed with respect to the average momenta $\bar{k}_1$ and $\bar{k}_2$ of the active partons.
For the active partons we can now introduce the momentum fractions
\be
x_1\ :=\ \frac{k_1^+}{p^+}\qquad\text{and}\qquad x_1^\prime\ :=\ \frac{k_1^{\prime +}}{p^{\prime +}}\,.
\ee
For later convenience we also introduce the average fraction
\be
\xb_1 = \frac{k_1^+ + k_1^{\prime +}}{p^+ + p^{\prime +}} = \frac{\bar{k}_1^+}{\bar{p}^{\,+}}\,,
\label{xbar-1-def}
\ee
which is related to $x_1$ and $x_1^\prime$ by
\be
x_1\= \frac{\xb_1+\xi}{1+\xi} \qquad\text{and}\qquad x_1^\prime\= \frac{\xb_1-\xi}{1-\xi}\,,
\label{individual-fractions}
\ee
respectively.
As for the processes in Refs.~\cite{Goritschnig:2009sq, DFJK1},
due to the large intrinsic scale given by the heavy quark mass $m_c$,
the transverse and minus (plus) components of the active \hbox{(anti)}parton momenta in the hard scattering kernel $\Tilde{H}$ are small as
compared to their plus (minus) components. Thus, the parton
momenta can be replaced by vectors lying in the scattering plane
formed by the parent hadron momenta.
For this assertion one only has to make the physically plausible assumptions
that the momenta are almost on mass-shell and
that their intrinsic transverse components
[divided by the respective momentum fractions (\ref{individual-fractions})]
are smaller than a typical hadronic scale of the order of $1$ GeV.
We thus make the following replacements:
\ba
k_1        &\to&   \left[\;{k}_1^+\,,\, \frac{x_1^2 \vdd}{8{k}_1^+}\,,
                     - x_1\frac{\vd}{2}\right]
                   \quad\text{with}\quad {k}_1^+ = x_1 p^+  \,, \nn\\[0.1em]
k^\prime_1 &\to&  \left[\;{k}_1^{\prime +},
                      \frac{m_c^2+x_1^{\prime 2}\vdd/4}
                      {2{k}_1^{\prime +}},  x_1^\prime\frac{\vd}{2} \right]
                   \quad\text{with}\quad {k}_1^{\prime +} = x_1^\prime p^{\prime +}  \,, \nn\\[0.1em]
k_2        &\to&   \left[\;\frac{x_2^2\vdd}{8{k}_2^-}\,,
                    \,{k}_2^-\,, \phantom{-} x_2\frac{\vd}{2} \right]
                   \quad\text{with}\quad {k}_2^- = x_2 q^-  \,, \nn\\[0.1em]
k^\prime_2 &\to&  \left[\frac{m_c^2+x_2^{\prime 2}\vdd/4}{2{k}_2^{\prime -}},
                                           {k}_2^{\prime -}, - x_2^\prime\frac{\vd}{2}\right]
                   \quad\text{with}\quad {k}_2^{\prime -} = x_2^\prime q^{\prime -}  \,.
\label{parton-mom}
\ea
As a consequence of these replacements it is then possible to explicitly perform the integrations
over $\bar{k}_1^-$, $\bar{k}_2^+$, $\bar{\VEC{k}}_{\perp 1}$ and $\bar{\VEC{k}}_{\perp 2}$.
Furthermore, the relative distance between the
\hbox{(anti-)}$S[ud]$-diquark and the \hbox{(anti-)}$c$-quark
field operators in the hadronic matrix elements is forced to be lightlike,
i.e., they have to lie on the light cone and thus the
time ordering of the field operators can be dropped.
After these simplifications one arrives at the following expression
for the $\ourprocess$ amplitude:
\begin{equation}
\begin{split}
M_{\mu \nu} =
&    \sum_{\rm a_i^{(\prime)},\, \alpha_i^{(\prime)}} \,
     \int d \bar{k}_1^{\,+} \theta(\bar{k}_1^{\,+}) \int \frac{d z_1^-}{2\pi} e^{i \bar{k}_1^{\,+} z_1^-}
     \int d \bar{k}_2^{\,-} \theta(\bar{k}_2^{\,-}) \int \frac{d z_2^+}{2\pi} e^{i \bar{k}_2^{\,-} z_2^+} \\
& \times \langle \overline{D^0} : p^\prime |
             \Psi^c_{a^\prime_1\alpha^\prime_1}(-\bar{z}_1/2)
             \Phi^{S[ud]}_{a_1}(+\bar{z}_1/2)
	 |p : p , \mu\rangle
     \,\, \Tilde{H}_{a_i^{(\prime)} \alpha^{\prime}_i}
           \left( \bar{k}_1 , \bar{k}_2 \right) \\
& \times \langle D^0 : q^\prime |
              \Phi^{S[ud]\dagger}_{a_2}(+\bar{z}_2/2)
              \overline{\Psi}^c_{a^\prime_2\alpha^\prime_2}(-\bar{z}_2/2)
	  |\bar{p} : q , \nu\rangle \,. \\
\label{eq:integration-1}
\end{split}
\end{equation}
From now on we will omit the color and spinor labels whenever this does not lead to ambiguities and
replace the field-operator arguments $\bar{z}_1$ and $\bar{z}_2$ by
their non-vanishing components $z_1^-$ and $z_2^+$, respectively.
Furthermore, if one uses $\bar{k}_1^{\,+} = \bar{x}_1 \bar{p}^{\,+}$ and
$\bar{k}_2^{\,-} = \bar{x}_2 \bar{q}^{\,-}$ to rewrite the
$\bar{k}_1^{\,+}$ and $\bar{k}_2^{\,-}$ integrations in the amplitude (\ref{eq:integration-1})
as integrations over the longitudinal momentum fractions $\bar{x}_1$ and $\bar{x}_2$, respectively,
one arrives at,
%
\be
\begin{split}
M_{\mu \nu} =
&\int d \bar{x}_1 \, \bar{p}^{\,+} \, \int \frac{d z_1^-}{2\pi} e^{i \bar{x}_1 \bar{p}^{\,+} z_1^-}
 \int d \bar{x}_2 \, \bar{q}^{\,-} \, \int \frac{d z_2^+}{2\pi} e^{i \bar{x}_2 \bar{q}^{\,-} z_2^+} \\
& \times \langle \overline{D^0} : p^\prime |
             \Psi^c(-z_1^-/2)
             \Phi^{S[ud]}(+z_1^-/2)
	 |p : p , \mu\rangle
     \,\, \Tilde{H}\left( \bar{x}_1\bar{p}^{\,+}, \bar{x}_2\bar{q}^{\,-} \right) \\	
& \times \langle D^0 : q^\prime |
             \Phi^{S[ud]\dagger}(+z_2^+/2)
             \overline{\Psi}^c(-z_2^+/2)
	  |\bar{p} : q , \nu\rangle \,.
\label{eq:integration}
\end{split}
\ee
%
As in Ref.~\cite{Goritschnig:2009sq} for $p \,\to\, \Lambda_c^+$ ($\bar{p} \,\to\, \overline{\Lambda}_c^-$),
the $p \,\to\, \overline{D^0}$ ($\bar{p} \,\to\, D^0$) transition matrix element is expected
to exhibit a pronounced peak with respect to the momentum fraction.
The position of the peak is approximately at
\be
x_0 \= \frac{m_c}{M} \= 0.68\,.
\ee
From Eq.~(\ref{partonic-threshold}) one then infers
that the relevant average momentum fractions
$\bar{x}_1$ and $\bar{x}_2$
have to be larger than the skewness $\xi$.
This means that the convolution integrals in Eq.~(\ref{eq:integration}) have to be
performed only from $\xi$ to 1 and not from 0 to 1.

In the following section we will analyze the soft hadronic matrix elements in some more detail.

\section{Hadronic Transition Matrix Elements}
\label{sec:quark-field-product}
%

%
Compared to Eq.~(\ref{ampl}) the Fourier transforms of the hadronic matrix elements
for the $p \to \overline{D^0}$ and $\bar{p} \to D^0$ transitions
are rendered to Fourier integrals solely over $z_1^-$ and $z_2^+$, respectively.
Hence we have to study the integral
%
\be
\bar{p}^{\,+} \, \int \frac{d z_1^-}{2\pi} e^{i \bar{x}_1\bar{p}^{+} z_1^-}
\langle \overline{D^0} : p^\prime |  \Psi^c(-z_1^-/2)
             \Phi^{S[ud]}(+z_1^-/2)|p : p , \mu\rangle \,,
\label{soft-hadronic-matrix-element}
\ee
%
over the $p \to \overline{D^0}$ transition matrix element and the integral
%
\be
\bar{q}^- \, \int \frac{d z_2^+}{2\pi} e^{i \bar{x}_2\bar{q}^{-} z_2^+}
\langle D^0 : q^\prime |  \Phi^{S[ud]\dagger}(+z_2^+/2)
                    \overline{\Psi}^c(-z_2^+/2)|\bar{p} : q , \nu\rangle \,,
\label{soft-anti-hadronic-matrix-element}
\ee
%
over the $\bar{p} \to D^0$ transition matrix element
instead of Eqs.~(\ref{soft-Tordered-hadronic-matrix-element}) and
(\ref{soft-Tordered-anti-hadronic-matrix-element}), respectively.
We will first concentrate on the
$p \to \overline{D^0}$ transition (\ref{soft-hadronic-matrix-element})
and investigate the product of field operators
$\Psi^c(-z^-_1/2)\Phi^{S[ud]}(+z^-_1/2)$.
For this purpose we consider the $c-$quark field operator $\Psi^c$
in the hadron frame of the outgoing $\overline{D^0}$,
cf. e.g., Refs.~\cite{Kogut:1969xa, Brodsky:1989pv},
where the $\overline{D^0}$ has no transverse momentum component.
It can be reached from our symmetric CMS by a transverse boost
\cite{Dirac:1949cp, Leutwyler:1977vy} with the boost parameters
\be
b^+ = \left(1-\xi\right) \bar{p}^{\,+}
\quad\text{and}\quad
\VEC{b}_\perp = \frac{\bm{\Delta}_\perp}{2}\,.
\ee
In this hadron-out frame we write the field operator in terms of its
\lq\lq good\rq\rq\ and \lq\lq bad\rq\rq\ light cone components,
\be
\Psi^c \= \frac12 (\gamma^-\gamma^+ +  \gamma^+\gamma^-) \, \Psi^c
       \,\equiv\, \Psi^{c}_+ + \Psi^{c}_-\,,
\ee
by means of the good and bad projection operators
$\mathcal{P}^+ = \frac12 \gamma^- \gamma^+$ and
$\mathcal{P}^- = \frac12 \gamma^+ \gamma^-$, respectively.
After doing that we eliminate the $\gamma^-$ appearing in
$\mathcal{P}^+$ and $\mathcal{P}^-$ by
using the $\bar{c}-$quark energy projector
\be
\sum_{\lambda^\prime_1} v({k}^\prime_1,\lambda^\prime_1)
                     \bar{v}({k}^\prime_1,\lambda^\prime_1)\=
                              {k}^\prime_1\cdot\gamma - m_c \,.
\ee
In the hadron-out frame it explicitely takes on the form
\be
\sum_{\lambda^\prime_1} v(\hat{k}^\prime_1,\lambda^\prime_1)
                     \bar{v}(\hat{k}^\prime_1,\lambda^\prime_1)\=
{k}_1^{\prime +}\gamma^- + \frac{m_c^2}{2{k}_1^{\prime +}}\gamma^+ - m_c \,,
\label{eq:proj-mass}
\ee
since there the $\bar{c}-$quark momentum is
\be
\hat{k}_1^\prime =
\left[
{k}^{\prime +}_1,
\frac{m_c^2}{2{k}_1^{\prime +}},\tr0
\right]\,.
\ee
With those replacements the $c-$quark field operator becomes
\ba
\Psi^c &=& \frac{1}{2{k}_1^{\prime +}} \sum_{\lambda^\prime_1} \Big\{
       v(\hat{k}^\prime_1,\lambda^\prime_1)
       \left(\bar{v}(\hat{k}^\prime_1,\lambda^\prime_1)\gamma^+\Psi^c\right) \nn\\
       &+& \gamma^+\Big[ v(\hat{k}^\prime_1,\lambda^\prime)
        (\bar{v}(\hat{k}^\prime_1,\lambda^\prime_1)\Psi^c)
                                            + 2m_c\Psi^c\Big]\Big\}\,.
\label{eq:decomp-mass}
\ea
As in the case of $p\bar{p} \,\to\, \Lambda_c^+\overline{\Lambda}_c^-$
in Ref.~\cite{Goritschnig:2009sq}, one can argue that the contribution coming from
$\left(\bar{v}(\hat{k}^\prime_1,\lambda^\prime_1)\gamma^+\Psi^c\right)$
dominates over the one in the square brackets, and thus the latter one can be neglected.
Since this dominant contribution can be considered as a plus component of a four-vector,
one can immediately boost back to our symmetric CMS where it then still holds that
\be
\Psi^c(-z_1/2) \= \frac{1}{2{k}_1^{\prime +}} \sum_{\lambda^\prime_1} 
                        v({k}^\prime_1,\lambda^\prime_1)
   \left(\bar{v}({k}^\prime_1,\lambda^\prime_1)\gamma^+\Psi^c(-z_1/2)\right)\,.
\label{eq:decomp-c}
\ee
Furthermore, one can even show that in
$\left( \bar{v}(k_1^\prime ,\,\lambda_1^\prime )\gamma^+\Psi^c(-z_1^-/2) \right)$
on the right-hand side of Eq.~(\ref{eq:decomp-c}) only the good component of
$\Psi^c(-z_1^-/2)$ is projected out, since
\be
  \bar{v}(k_1^\prime ,\, \lambda_1^\prime )\gamma^+\Psi^{c}(-z_1^-/2)
 =\bar{v}(k_1^\prime ,\, \lambda_1^\prime )\gamma^+ \mathcal{P}^+ \Psi^{c}(-z_1^-/2) \,.
\ee
Finally, we note that such manipulations are not necessary for the
scalar field operator $\Phi^{S[ud]}$ of the $S[ud]$ diquark.
Putting everything together gives for the $p \to \overline{D^0}$
transition matrix element (\ref{soft-hadronic-matrix-element})
%
\be
\begin{split}
 &\bar{p}^{\,+} \, \int \frac{d z_1^-}{2\pi} e^{i \bar{x}_1\bar{p}^{+} z_1^-}
 \langle \bar{D^0} : p^\prime |  \Psi^c(-z_1^-/2)
             \Phi^{S[ud]}(+z_1^-/2)|p : p , \mu\rangle
 \,=\, \\
 &\frac{\bar{p}^{\,+}}{2k_1^{\prime +}} \,\sum_{\lambda_1^\prime} \,\int \frac{d z_1^-}{2\pi}
 e^{i \bar{x}_1\bar{p}^{+} z_1^-}
 \langle \bar{D^0} : p^\prime |
             v\left(k_1^\prime ,\,\lambda_1^\prime \right)
	     \left( \bar{v}(k_1^\prime ,\,\lambda_1^\prime )\gamma^+\Psi^{c}_+(-z_1^-/2) \right)
             \Phi^{S[ud]}(+z_1^-/2)
 |p : p , \mu\rangle \,.
\label{explicit-hadronic-matrix-element}
\end{split}
\ee
%

%
Proceeding in an analogous way for the $\bar{p} \to D^0$ transition matrix element,
where the role of the $+$ and $-$ components are interchanged,
we get for Eq.~(\ref{soft-anti-hadronic-matrix-element})
%
\be
\begin{split}
&\bar{q}^- \, \int \frac{d z_2^+}{2\pi} e^{i \bar{x}_2\bar{q}^{-} z_2^+}
 \langle D^0 : q^\prime |  \Phi^{S[ud]\dagger}(+z_2^+/2)
                    \overline{\Psi}^c(-z_2^+/2)|\bar{p} : q , \nu\rangle
\,=\, \\
&\frac{\bar{q}^-}{2k_2^{\prime -}} \,\sum_{\lambda_2^\prime} \,\int \frac{d z_2^+}{2\pi}
 e^{i \bar{x}_2\bar{q}^{-} z_2^+}
 \langle D^0 : q^\prime |  \Phi^{S[ud]\dagger}(+z_2^+/2)
  \left( \overline{\Psi}^{c}_+(-z_2^+/2) \gamma^- u\left(k_2^\prime,\lambda_2^\prime\right) \right)
  \bar{u}\left(k_2^\prime,\lambda_2^\prime\right)
 |\bar{p} : q , \nu\rangle \,.
\label{explicit-anti-hadronic-matrix-element}
\end{split}
\ee
%
Also here only the good components of the quark field are projected out on the right-hand side.
Using now Eqs.~(\ref{explicit-hadronic-matrix-element})
and (\ref{explicit-anti-hadronic-matrix-element}) and
attaching the spinors $v\left(k_1^\prime ,\,\lambda_1^\prime \right)$
and $\bar{u}\left(k_2^\prime,\lambda_2^\prime\right)$ to the hard subprocess amplitude
$\Tilde{H}$ by introducing
\be
 H_{\lambda_1^\prime\,,\lambda_2^\prime}\left( \bar{x}_1 \,, \bar{x}_2\right) :=
  \bar{u}\left(k_2^\prime,\lambda_2^\prime\right)
  \Tilde{H}\left( \bar{x}_1\bar{p}^{\,+}, \bar{x}_2\bar{q}^{\,-} \right)
  v\left(k_1^\prime ,\,\lambda_1^\prime \right) \,,
\label{def:subprocess-amplitudes}
\ee
we get for the $\ourprocess$ amplitude (\ref{eq:integration})
\be
\begin{split}
M_{\mu \nu} =
 &\frac{1}{4(\bar{p}^{\,+})^2}
 \, \sum_{\lambda_1^\prime , \lambda_2^\prime}
 \, \int d \bar{x}_1 \, \int d \bar{x}_2 \,
 \, H_{\lambda_1^\prime\,,\lambda_2^\prime}\left( \bar{x}_1 \,, \bar{x}_2\right)
 \, \frac{1}{\bar{x}_1-\xi} \, \frac{1}{\bar{x}_2-\xi} \\
 & \times \bar{v}(k_1^\prime ,\,\lambda_1^\prime )\gamma^+ \quad
    \bar{p}^{\,+} \,\int \frac{d z_1^-}{2\pi} \, e^{i \bar{x}_1\bar{p}^{+} z_1^-} \,
    \langle \bar{D^0} : p^\prime |
    \Psi^c_+(-z_1^-/2)
    \Phi^{S[ud]}(+z_1^-/2)
    |p : p , \mu\rangle \\
 &\times \bar{q}^{\,-} \,\int \frac{d z_2^+}{2\pi} \, e^{i \bar{x}_2\bar{q}^{-} z_2^+} \,
    \langle D^0 : q^\prime |  \Phi^{S[ud]\dagger}(+z_2^+/2)
    \overline{\Psi}^c_+(-z_2^+/2)
    |\bar{p} : q , \nu\rangle \quad
    \gamma^- u\left(k_2^\prime,\lambda_2^\prime\right)\,.
\label{Mmunu}
\end{split}
\ee
Introducing the abbreviations
\be
\mathcal{H}_{\lambda_1^\prime \mu}^{\bar{c}S} \,:=\,
\bar{v}(k_1^\prime ,\,\lambda_1^\prime )\gamma^+ \,\,
    \bar{p}^{\,+} \,\int \frac{d z_1^-}{2\pi} \, e^{i \bar{x}_1\bar{p}^{+} z_1^-} \,
    \langle \bar{D^0} : p^\prime |
    \Psi^c_+(-z_1^-/2)
    \Phi^{S[ud]}(+z_1^-/2)
    |p : p , \mu\rangle
\label{eq:mathcalH}
\ee
and
\be
\mathcal{H}_{\lambda_2^\prime \nu}^{c\bar{S}} \,:=\,
\bar{q}^{\,-} \,\int \frac{d z_2^+}{2\pi} \, e^{i \bar{x}_2\bar{q}^{-} z_2^+} \,
    \langle D^0 : q^\prime |  \Phi^{S[ud]\dagger}(+z_2^+/2)
    \overline{\Psi}^c_+(-z_2^+/2)
    |\bar{p} : q , \nu\rangle \,\,
    \gamma^- u\left(k_2^\prime,\lambda_2^\prime\right) \,,
\label{eq:mathcalantiH}
\ee
for the pertinent projections of the hadronic transition matrix elements,
we can write the hadronic scattering amplitude in a more compact form:
\be
M_{\mu \nu} =
 \frac{1}{4(\bar{p}^{\,+})^2}
 \, \sum_{\lambda_1^\prime , \lambda_2^\prime}
 \, \int d \bar{x}_1 \, \int d \bar{x}_2 \,
 \, H_{\lambda_1^\prime\,,\lambda_2^\prime}\left( \bar{x}_1 \,, \bar{x}_2\right)
 \, \frac{1}{\bar{x}_1-\xi} \, \frac{1}{\bar{x}_2-\xi}
 \,\mathcal{H}_{\lambda_1^\prime \mu}^{\bar{c}S}
 \,\mathcal{H}_{\lambda_2^\prime \nu}^{c\bar{S}}\,.
\label{Mmunu-short}
\ee

\section{Overlap Representation of $\mathcal{H}_{\lambda_1^\prime \mu}^{\bar{c}S}$}
\label{sec:overlap-representation}
%

In the following section we will derive a representation for the hadronic
$p \,\to\, \overline{D^0}$ and $\bar{p} \,\to\, D^0$ transition matrix elements
as an overlap of hadronic light-cone wave functions (LCWFs) for the
valence Fock components of $p$ and $\overline{D^0}$ \cite{Diehl:2000xz}.
Since we only need them for $\bar{x} > \xi$, i.e., 
in the Dokshitzer-Gribov-Lipatov-Altarelli-Parisi region,
the hadronic transition matrix elements admit such a representation.
For doing that we will make use of the Fock expansion of the hadron states and
the Fourier decomposition of the partonic field operators in light-cone quantum field theory.

At a given light-cone time, say $z^+ = 0$,
the good independent dynamical field components 
$\Phi^{S[ud]}$ and $\Psi^c_+\left(-z_1^-/2\right)$
of the $S[ud]$ diquark and the $c$ quark, respectively, have the Fourier decomposition
\be
\begin{split}
\Phi^{S[ud]}\left(+z_1^-/2\right) \=
 \int\frac{dk_1^+}{k_1^+}\,\int\frac{d^2 k_{1\perp}}{16\pi^3}\,\theta\left(k_1^+\right)
  &\left[
  a\left(S[ud]:\,k_1^+,\,\VEC{k}_{1\perp}\right)
  e^{-\imath k_1^+ \frac{z_1^-}{2}} \right. \\
  &\left. +
  b^\dagger\left(S[ud]:\,k_1^+,\,\VEC{k}_{1\perp}\right)
  e^{+\imath k_1^+ \frac{z_1^-}{2}}
 \right]
\label{Sud-field-operator}
\end{split}
\ee
and
\be
\begin{split}
\Psi^{c}_+\left(-z_1^-/2\right) \=
 \int\frac{dk_1^{\prime +}}{k_1^{\prime +}}\,
 \int\frac{d^2 k_{1\perp}^\prime}{16\pi^3}\,
 \theta\left(k_1^{\prime +}\right)
 \sum_{\lambda_1^\prime}
  &\left[
  c\left(c:\,k_1^{\prime +},\,\VEC{k}_{1\perp}^\prime,\,\lambda_1^\prime\right)
  u_+\left(k_1^\prime,\,\lambda_1^\prime\right)
  e^{+\imath k_1^{\prime +} \frac{z_1^-}{2}} \right. \\
  &\left. +
  d^\dagger\left(c:\,k_1^{\prime +},\,\VEC{k}_{1\perp}^\prime,\,\lambda_1^\prime\right)
  v_+\left(k_1^\prime,\,\lambda_1^\prime\right)
  e^{-\imath k_1^{\prime +} \frac{z_1^-}{2}}
 \right] \,.
\label{c-field-operator}
\end{split}
\ee
The spinors $u_+$ and $v_+$ are the good components of the (anti)quark
spinors $u$ and $v$, i.e., $u_+ \= \mathcal{P}^+ u$ and $v_+ \= \mathcal{P}^+ v$.
The operators $a$ and $b^\dagger$ are the annihilator of an $S[ud]$ diquark and 
the creator of  an $\overline{S[ud]}$ diquark, respectively.
The operator $c$ annihilates a $c$ quark and
the operator $d^\dagger$ creates a $\bar{c}$ quark.
Their action on the vacuum gives the single-parton states
\ba
a^\dagger\left(S[ud]:\,k_1^+,\,\VEC{k}_{1\perp}\right) \mid 0 \,\rangle
 &=& \mid S[ud]:\,k_1^+,\,\VEC{k}_{1\perp} \rangle \,, \\
b^\dagger\left(S[ud]:\,k_1^+,\,\VEC{k}_{1\perp}\right)\mid 0 \,\rangle
 &=& \mid \overline{S[ud]}:\,k_1^+,\,\VEC{k}_{1\perp} \rangle \,, \\
c^\dagger\left(c:\,k_1^{\prime +},\,\VEC{k}_{1\perp}^\prime,\,\lambda_1^\prime\right)\mid 0 \,\rangle
 &=& \mid c:\,k_1^{\prime +},\,\VEC{k}_{1\perp}^\prime,\,\lambda_1^\prime \rangle \,, \\
d^\dagger\left(c:\,k_1^{\prime +},\,\VEC{k}_{1\perp}^\prime,\,\lambda_1^\prime\right)\mid 0 \,\rangle
 &=& \mid \bar{c}:\,k_1^{\prime +},\,\VEC{k}_{1\perp}^\prime,\,\lambda_1^\prime \rangle \,,
\ea
which are normalized as follows: 
\be
 \langle  \cdots :\,k^{\prime +},\,\VEC{k}^\prime_\perp,\,\lambda^\prime
 \mid
 \cdots:\,k^{+},\,\VEC{k}_\perp,\,\lambda \rangle \=
 16\pi^3 \, k^+ \,
 \delta \left( k^{\prime +} - k^+ \right)
 \delta^{(2)} \left( \VEC{k}^\prime_\perp - \VEC{k}_\perp \right) \,
 \delta_{\lambda^\prime,\,\lambda} \,.
\ee
In the case of $S[ud]$ states no $\lambda^{(\prime)}$ and no $\delta_{\lambda^\prime,\,\lambda}$ appear.
This normalization is in accordance with the \hbox{(anti)}commutation relations
\be
\begin{split}
 \left[
  a\left(S[ud]:\,k^{\prime +},\,\VEC{k}^\prime_\perp\right),\,
  a^\dagger\left(S[ud]:\,k^+,\,\VEC{k}_\perp\right)
 \right]
 & \=
 \left[
  b\left(S[ud]:\,k^{\prime +},\,\VEC{k}^\prime_\perp\right),\,
  b^\dagger\left(S[ud]:\,k^+,\,\VEC{k}_\perp\right)
 \right] \\
  & \=
  16\pi^3 \, k^+ \,
  \delta \left( k^{\prime +} - k^+ \right)
  \delta^{(2)} \left( \VEC{k}^\prime_\perp - \VEC{k}_\perp \right)
\end{split}
\ee
and
\be
\begin{split}
 \{
  c\left(c:\,k^{\prime +},\,\VEC{k}^\prime_\perp,\,\lambda^\prime\right),\,
  c^\dagger\left(c:\,k^+,\,\VEC{k}_\perp,\,\lambda\right)
 \}
 & \=
 \{
  d\left(c:\,k^{\prime +},\,\VEC{k}^\prime_\perp,\,\lambda^\prime\right),\,
  d^\dagger\left(c:\,k^+,\,\VEC{k}_\perp,\,\lambda\right)
 \} \\
 & \=
  16\pi^3 \, k^+ \,
  \delta \left( k^{\prime +} - k^+ \right)
  \delta^{(2)} \left( \VEC{k}^\prime_\perp - \VEC{k}_\perp \right)
  \delta_{\lambda^\prime,\,\lambda} \,.
\end{split}
\ee

In the Fock state decomposition hadrons on the light front are replaced
by a superposition of parton states.
Taking only into account the valence Fock state,
the proton and the $\overline{D^0}$ state in our quark-diquark picture
are represented as
\be
\begin{split}
\mid p : p,\,\mu \rangle =
&\int d\tilde{x}\,\frac{d^2\tilde{k}_\perp}{16\pi^3} \,
 \psi_p\left(\tilde{x},\,\VEC{\tilde{k}}_{\perp}\right) \,
 \frac{1}{\sqrt{\tilde{x}(1-\tilde{x})}} \\
&\times\mid S_{\left[ud\right]}:\,\tilde{x}p^+,\,\VEC{\tilde{k}}_{\perp}+\tilde{x}\VEC{p}_{\perp} \rangle
 \mid u:\,(1-\tilde{x})p^+,\,-\VEC{\tilde{k}}_{\perp}+(1-\tilde{x})\VEC{p}_{\perp},\,\mu \rangle
\label{Eq_qD_ProtonState}
\end{split}
\ee
and
\be
\begin{split}
\mid \overline{D^0} : p^\prime \rangle =
&\int d\hat{x}^\prime\,\frac{d^2 \hat{k}^\prime_\perp}{16\pi^3}\,
 \psi_D\left( \hat{x}^\prime ,\,\VEC{\hat{k}^\prime}_{\perp} \right) \,
 \frac{1}{\sqrt{\hat{x}^\prime(1-\hat{x}^\prime)}} \,
 \frac{1}{\sqrt{2}}\sum_{\lambda^\prime} \left(2\lambda^\prime\right) \\
&\times\mid \bar{c}:\,\hat{x}^\prime p^{\prime +},\,\VEC{\hat{k}^\prime}_{\perp}+\hat{x}^\prime\VEC{p}^\prime_{\perp},\,\lambda^\prime \rangle\,
 \mid u:\,(1-\hat{x}^\prime)p^{\prime +} ,\,
 -\VEC{\hat{k}^\prime}_{\perp}+(1-\hat{x}^\prime)\VEC{p}^\prime_{\perp},\,-\lambda^\prime \rangle \,,
\label{Eq_qD_DState}
\end{split}
\ee
respectively, with normalization
\be
 \langle \cdots :\,p^{\prime +},\,\VEC{p}^\prime_\perp\,(,\,\lambda^\prime)
 \mid
 \cdots :\,p^{+},\,\VEC{p}_\perp\,(,\,\lambda) \rangle \=
 16\pi^3 \, p^+ \,
 \delta \left( p^{\prime +} - p^+ \right)
 \delta^{(2)} \left( \VEC{p}^\prime_\perp - \VEC{p}_\perp \right)
 \left(\delta_{\lambda^\prime,\,\lambda}\right) \,.
\ee
Also here, in the case of the pseudoscalar $D^0$ and $\overline{D^0}$ states
no $\lambda^{(\prime)}$ and no $\delta_{\lambda^\prime,\,\lambda}$ appear.
$\psi_p$ and $\psi_D$ are the LCWFs
of the proton and the $\overline{D^0}$, respectively,
which will be specified in Sec.~\ref{sec:Modelling-the-TDAs}.
The LCWFs do not depend on the total momentum of the hadron,
but only on the momentum coordinates of the partons relative to the hadron momentum.
Those relative momenta are most easily identified in the hadron frame of the parent hadron.
Here we have assumed that the partons inside the proton and the $\overline{D^0}$
have zero orbital angular momentum.
The arguments of the LCWFs are related to the
average momenta and momentum fractions by:
\be
\tilde{x} \= \frac{\bar{x} + \xi}{1 + \xi}\,,\quad
\tilde{{\bm k}}_\perp \= \bar{{\bm k}}_\perp - \frac{1-\bar{x}}{1 + \xi}\frac{{\bm \Delta}_\perp}{2} \,,
\ee
\be
\hat{x}^\prime \= \frac{\bar{x} - \xi}{1 - \xi}\,,\quad
\hat{{\bm k}}_\perp^\prime \= \bar{{\bm k}}_\perp + \frac{1-\bar{x}}{1 - \xi}\frac{{\bm \Delta}_\perp}{2} \,.
\ee
Using the expressions above we can write the hadronic matrix elements
appearing in Eq.~(\ref{Mmunu}) as
\be
\begin{split}
\mathcal{H}_{\lambda_1^\prime \mu}^{\bar{c}S}
\=& - 2\sqrt{2}\mu\bar{p}^{\,+} \, \int\,\frac{d\bar{x}d^2\bar{k}_\perp}{16\pi^3}
    \, \sqrt{\frac{\bar{x}-\xi}{\bar{x}+\xi}}
    \, \psi_D
    \left(\hat{x}^\prime(\bar{x},\,\xi),\,\hat{\VEC{k}}^\prime_\perp(\bar{\VEC{k}}_\perp\,,\bar{x}\,,\xi)\right) \\
&\phantom{=}\, \times \,
    \psi_p
    \left(\tilde{x}(\bar{x},\,\xi),\,\tilde{\VEC{k}}_\perp(\bar{\VEC{k}}_\perp\,,\bar{x}\,,\xi)\right)
    \, \delta\left(\bar{x}_1 - \bar{x}\right)\delta_{-\lambda_1^\prime ,\, \mu} \,,
\label{hadronic-transition-overlap}
\end{split}
\ee
for the $p \,\to\, \overline{D^0}$ transition and
\be
\begin{split}
\mathcal{H}_{\lambda_2^\prime \mu}^{c\bar{S}}
\=& - 2\sqrt{2}\nu\bar{q}^{\,-}\, \int\,\frac{d\bar{y}d^2\bar{l}_\perp}{16\pi^3}
    \, \sqrt{\frac{\bar{y}-\xi}{\bar{y}+\xi}}
    \, \psi_D
    \left(\hat{y}^\prime(\bar{y},\,\xi),\,\hat{\VEC{l}}^\prime_\perp(\bar{\VEC{l}}_\perp\,,\bar{y}\,,\xi)\right) \\
&\phantom{=}\, \times   \,
    \psi_p
    \left(\tilde{y}(\bar{y},\,\xi),\,\tilde{\VEC{l}}_\perp(\bar{\VEC{l}}_\perp\,,\bar{y}\,,\xi)\right)
    \, \delta\left(\bar{x}_2 - \bar{y}\right)\delta_{-\lambda_2^\prime ,\, \nu} \,,
\label{anti-hadronic-transition-overlap}
\end{split}
\ee
for the $\bar{p} \,\to\, D^0$ transition.
Here we have used that
\be
 \bar{v}\left(k_1^\prime,\,\lambda_1^\prime\right) \gamma^+ v\left(k_1^\prime,\,-\mu\right)
 \= 2 k_1^{\prime\,+}
\quad\text{and}\quad
 \bar{u}\left(k_2^\prime,\,-\nu\right) \gamma^- u\left(k_2^\prime,\,\lambda_2^\prime\right)
 \= 2 k_2^{\prime\,-} \,.
\label{eq:vbar-gammaplus-v-etc}
\ee
Collecting all pieces we finally get
\be
\begin{split}
M_{\mu \nu} \=
 & 2\mu\nu
 \, \int d \bar{x}_1 \, \int d \bar{x}_2 \,
 \, H_{-\mu,\,-\nu}\left( \bar{x}_1 \,, \bar{x}_2\right)
 \, \frac{1}{\sqrt{\bar{x}_1^2-\xi^2}} \, \frac{1}{\sqrt{\bar{x}_2^2-\xi^2}} \\
 & \times \, \int\,\frac{d^2\bar{k}_\perp}{16\pi^3} \,
\, \psi_D
\left(\hat{x}^\prime(\bar{x}_1,\,\xi),\,\hat{\VEC{k}}^\prime_\perp(\bar{\VEC{k}}_\perp\,,\bar{x}_1\,,\xi)\right)
    \psi_p
    \left(\tilde{x}(\bar{x}_1,\,\xi),\,\tilde{\VEC{k}}_\perp(\bar{\VEC{k}}_\perp\,,\bar{x}_1\,,\xi)\right) \\
 & \times \, \int\,\frac{d^2\bar{l}_\perp}{16\pi^3} \,
\, \psi_D
\left(\hat{y}^\prime(\bar{x}_2,\,\xi),\,\hat{\VEC{l}}^\prime_\perp(\bar{\VEC{l}}_\perp\,,\bar{x}_2\,,\xi)\right)
    \psi_p
    \left(\tilde{y}(\bar{x}_2,\,\xi),\,\tilde{\VEC{l}}_\perp(\bar{\VEC{l}}_\perp\,,\bar{x}_2\,,\xi)\right) \,.
\label{Mmunu-overlap-final}
\end{split}
\ee

Furthermore, we can take advantage of the expected shape of the
$p \,\to\, \overline{D^0}$ ($\bar{p} \,\to\, D^0$) transition matrix elements.
Due to their pronounced peak around $x_0$ only kinematical regions
in the hard scattering amplitude close to the peak position are
enhanced by the hadronic transition matrix elements.
For the hard partonic subprocess we, therefore, apply a
\lq\lq peaking approximation\rq\rq, i.e., we replace the momentum fractions
appearing in the hard-scattering amplitude by $x_0$.
Then the hard scattering amplitude can be pulled out of the convolution integral and the $\ourprocess$ amplitude simplifies further to
\be
\begin{split}
M_{\mu \nu} \,=\,
& 2\mu\nu 
  \, H_{-\mu,\,-\nu}\left( x_0,\, x_0\right)
  \, \Big[
  \, \int_\xi^1 d \bar{x} \,
  \, \frac{1}{\sqrt{\bar{x}^2-\xi^2}}
  \, \int\,\frac{d^2\bar{k}_\perp}{16\pi^3} \\
&  \,
 \psi_D
 \left(\hat{x}^\prime(\bar{x},\,\xi),\,\hat{\VEC{k}}^\prime_\perp(\bar{\VEC{k}}_\perp\,,\bar{x}\,,\xi)\right)
 \psi_p
 \left(\tilde{x}(\bar{x},\,\xi),\,\tilde{\VEC{k}}_\perp(\bar{\VEC{k}}_\perp\,,\bar{x}\,,\xi)\right)
 \, \Big]^2 \,,
\label{Mmunu-overlap-final-peaking-approx}
\end{split}
\ee
where the term in square bracket can be considered as a sort of generalized form factor.

\section{Hard Scattering Subprocess}
\label{sec:hard-subprocess}
%

%
Before we start to specify the LCWFs occuring in this overlap representation
of the $p \,\to\, \overline{D^0}$ ($\bar{p} \,\to\, D^0$) transition,
we will first calculate scattering amplitudes of the hard partonic
$S[ud]\,\overline{S[ud]} \,\to\, \bar{c}c$ subprocess within the
peaking approximation.
\begin{figure}[h]
\begin{center}
\includegraphics[width=.70\textwidth, clip=true]{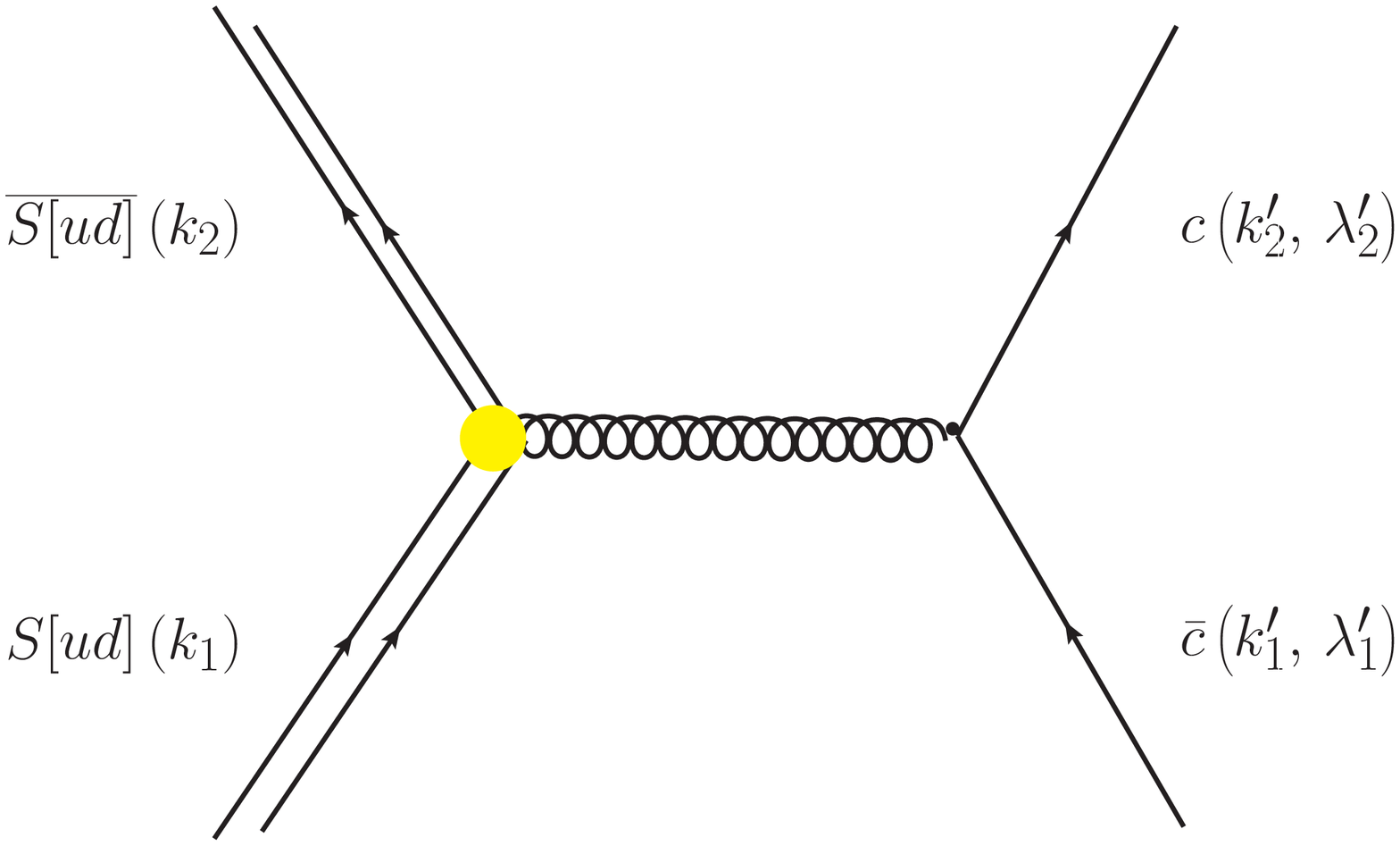}
\end{center}
\caption{The hard scattering process on the partonic level
$S[ud]\overline{S[ud]} \,\to\, c\bar{c}$.}
\label{fig:hard-subprocess}
\end{figure}
The hard-scattering amplitudes for the hard partonic subprocess,
as shown in Fig.~\ref{fig:hard-subprocess}, is given by
\be
H_{\lambda_1^\prime,\lambda_2^\prime} \,=\,
 \imath \frac{4}{9} \left(
  -\imath g_s \bar{u}\left(k_2^\prime,\lambda_2^\prime\right)
  \gamma^\mu v\left(k_1^\prime,\lambda_1^\prime\right)
 \right)
 \frac{-\imath g_{\mu\nu}}{\left( k_1 + k_2 \right)^2}
 \left(
 \left(-\imath g_s F_s \right) \left(k_1 - k_2\right)^\nu
 \right) \,.
\label{hard-amplitude}
\ee
$4/9$ is the color factor which we have attached to the hard-scattering amplitude.
$g_s=\sqrt{4\pi\alpha_s}$ is the \lq\lq usual\rq\rq\ strong coupling constant
and $F_s$ denotes the diquark form factor at the gluon-diquark vertex.
This diquark form factor takes care of the composite nature of the $S[ud]$ diquark
and the fact that for large $s$ the diquark should dissolve into quarks.
We have taken the phenomenological form factor from Ref.~\cite{Kroll:1993zx}, namely,
\be
 F_s(\hat{s}) \=
 \mid \frac{ Q_0^2 }{ Q_0^2 - \hat{s} } \mid \,,\quad
 Q_0^2 \= 3.22 \gev^2 \,,\quad
 \hat{s} > Q_0^2 \,.
\label{eq:F_s}
\ee
It is just the analytic continuation of a spacelike form factor
to the timelike region. The original spacelike form factor was introduced in  Ref.~\cite{Anselmino:1987vk} 
(where it has been obtained from fits to the structure functions of deep inelastic lepton-hadron scattering,
to the electromagnetic proton form factor and elastic proton-proton data at large momentum transfer). It should be remarked here that
such a continuation is not unique; the form factor can acquire unknown phases when doing the continuation.
But, fortunately, such phases are irrelevant with respect to the physics, which is the reason for
taking the absolute value in (\ref{eq:F_s}).

With the help of the peaking approximation we can express the subprocess amplitudes in terms of the kinematical variables of the full process. For the different helicity combinations we explicitely have,
\ba
 H_{++} &\,=\,&  + 4\pi\alpha_s(x_0^2 s) \, F_s(x_0^2 s)
 \,\frac{4}{9}\,  \frac{2M}{\sqrt{s}} \cos\theta \,, \nn\\
%
 H_{+-} &\,=\,&  - 4\pi\alpha_s(x_0^2 s) \, F_s(x_0^2 s)
 \,\frac{4}{9}\,  \sin\theta \,, \nn\\
%
 H_{-+} &\,=\,&  - 4\pi\alpha_s(x_0^2 s) \, F_s(x_0^2 s)
 \,\frac{4}{9}\,  \sin\theta \,, \nn\\
%
 H_{--} &\,=\,&  - 4\pi\alpha_s(x_0^2 s) \, F_s(x_0^2 s)
 \,\frac{4}{9}\,  \frac{2M}{\sqrt{s}} \cos\theta \,.
\label{eq:subprocess-amplitudes}
\ea
%

\section{Modelling the Hadronic Transition Matrix Elements}
\label{sec:Modelling-the-TDAs}
%

In order to make numerical predictions we have to specify the LCWFs for the proton and the $D^0$.
We will use wave functions of the form
\be
 \psi \sim e^{-a^2 \sum_i \frac{\VEC{k}_{i\perp}^2+m_i^2}{x_i}} \,,
\label{eq:oszi-LCWF}
\ee
which can be traced back to a harmonic oscillator ansatz \cite{Wirbel:1985ji}
that is transformed to the light cone \cite{Guo:1991eb}.
In Ref.~\cite{Korner:1991zxKorner:1992uw} it was adapted to the case
of baryons within a quark-diquark picture.
According to Refs.~\cite{Goritschnig:2009sq, Kroll:1988cd} we write the wave functions
of a proton in an $S[ud]u$ Fock state as
\be
 \psi_p(x,\,\VEC{k}_\perp) \=
 N_p \, x \, e^{-a_p^2 \frac{\VEC{k}_\perp^2}{x(1-x)}}
\label{eq:proton-LCWF}
\ee
and the one of a pseudoscalar $D^0$ meson in a $u\bar{c}$  Fock state as
\be
 \psi_D(x,\,\VEC{k}_\perp) \=
 N_D \, e^{-a_D^2 M^2 \frac{(x-x_0)^2}{x(1-x)}} \, e^{-a_D^2 \frac{\VEC{k}_\perp^2}{x(1-x)}} \,.
\label{eq:D-LCWF}
\ee
Here $x$ is the momentum fraction of the active constituent,
the $S[ud]$ diquark or the $\bar{c}$ quark, respectively.
The mass exponential in Eq.~(\ref{eq:D-LCWF}) generates the expected pronounced peak at $x \approx x_0$
and is a slightly modified version of the one given in Ref.~\cite{Korner:1991zxKorner:1992uw}.

In each of the wave functions, Eqs.~(\ref{eq:proton-LCWF}) and (\ref{eq:D-LCWF}), we have two free parameters:
on the one hand the transverse size parameter $a_{p/D}$ and, on the other hand,
the normalization constant $N_{p/D}$.
The parameters can be associated with the mean intrinsic transverse momentum squared
$\langle \VEC{k}_\perp^2 \rangle_{p/D}$ of the active constituent inside its parent hadron
and with the probability to find the hadron in the specific Fock state
(or with the decay constant $f_{p/D}$ of the corresponding hadron).
The probabilities and the intrinsic transverse momenta for the valence Fock states
as given in Eqs.~(\ref{Eq_qD_ProtonState}) and (\ref{Eq_qD_DState})
can be calculated as
\be
 P_{p/D} \= \int dx \int \frac{d^2 k_\perp}{16\pi^3} \mid \psi_{p/D}(x,\,\VEC{k}_\perp) \mid^2
\label{eq:probs}
\ee
and
\be
 \langle\VEC{k}_\perp^2\rangle_{p/D} \= \frac{1}{P_{p/D}}
  \int dx \int \frac{d^2 k_\perp}{16\pi^3}
  \VEC{k}_\perp^2 \mid \psi_{p/D}(x,\,\VEC{k}_\perp) \mid^2 \,,
\label{eq:intr-transv-mom}
\ee
respectively.
Inserting the wave functions (\ref{eq:proton-LCWF}) and (\ref{eq:D-LCWF})
into Eqs.~(\ref{eq:probs}) and (\ref{eq:intr-transv-mom}), we obtain
\be
P_p \=
\frac{N_p^2}{640\pi^2a_p^2}
\,,\qquad
\langle \VEC{k}_{\perp}^{2} \rangle_p \=
\frac{2}{21a_p^2}
\ee
and
\be
P_D \=
\frac{N_D^2}{32 \pi^2 a_D^2} I_{11}(a_D^2)
\,,\qquad
\langle \VEC{k}_{\perp}^{2} \rangle_D \=
\frac{1}{2 a_D^2}\frac{I_{22}(a_D^2)}{I_{11}(a_D^2)}\,,
\ee
where we have introduced the abbreviation
\be
I_{nm}(a_D^2)\,:=\,\int_0^1 dx \, x^n \left(1-x\right)^m \exp{\left[-2 a_D^2 M^2 \frac{(x-x_0)^2}{x(1-x)}\right]} \,.
\ee
For the proton we use the same parameters as in Refs.~\cite{Goritschnig:2009sq, Kroll:1988cd}.
We choose $a_p = 1.1 \text{GeV}^{-1}$ for the oscillator parameter
and $P_p = 0.5$ for the valence Fock state probability.
Choosing $P_p = 0.5$ for the proton may appear rather large at first sight.
As a bound state of two quarks a diquark embodies also gluons and sea quarks
and thus effectively incorporates also higher Fock states.
Therefore, a larger probability than one would expect for a 3-quark valence Fock state
and a larger transverse size of the quark-diquark state appear plausible.
Choosing the parameter values as stated above we get for the proton
\be
 \sqrt{\langle \VEC{k}_\perp^2\rangle_p} = 280\,\text{MeV}
 \qquad\text{and}\qquad
 N_p = 61.818\,\gev^{-2} \,.
\ee

For the $D$ meson we fix the two parameters such that we get certain values
for the valence Fock state probability $P_D$ and the decay constant $f_D$.
The decay constant $f_D$ is defined by the relation
\be
 \langle 0 \mid
 \overline{\Psi}^u(0) \gamma^\mu \gamma_5 \Psi^c(0)
 \mid D^0:\,p \rangle
\= \imath f_D p^\mu \,.
\ee
Taking the plus component and inserting the fields as given in
Sec.~\ref{sec:overlap-representation}, we get (omitting phases)
\be
 2\sqrt{6} \int dx \frac{d^2 k_\perp}{16\pi^3} \psi_D (x,\,\vk)
 \= f_D \,,
\ee
such that
\be
 N_D \= \frac{16 \pi^2 a_D^2 f_D}{2 \sqrt{6} I_{11}(a_D^2/2)} \,.
\ee
As value for the decay constant we take the experimental value
$f_D = 206$ MeV from Ref.~\cite{PDG};
for the valence Fock state probability we choose $P_D = 0.9$.
This amounts to $a_D = 0.864 \text{GeV}^{-1}$.
As values for the normalization constant and for the root mean square of
the intrinsic transverse momentum of the active quark we then get
\be
 \sqrt{\langle \VEC{k}_\perp^2\rangle_D} = 383\,\text{MeV}
 \qquad\text{and}\qquad
 N_D = 55.2\,\gev^{-2} \,,
\ee
respectively.

Let us now turn to the issue of the error assesment with respect to the parameters.
For the decay constant of the $D^0$ meson we take $f_D = 206 \,\pm 8.9$ MeV
as stated in Ref.~\cite{PDG}. The valence Fock state probability
of the $D$ meson $P_D$ is varied between $0.8$ and $1$.
We do not take into account the uncertainties of the parameters
appearing in the proton LCWF.
They are small compared to the ones of the $D$ meson LCWF since
they have been determined from detailed studies of other processes.
The influence of the parameter uncertainties on the
cross sections are indicated by grey error bands
in Figs.~\ref{fig:dsdt} and \ref{fig:sigma}.

We now turn to the wave function overlap
as derived in Sec.~\ref{sec:overlap-representation}.
When taking the model wave functions (\ref{eq:proton-LCWF})
and (\ref{eq:D-LCWF}) we explicitely get
\be
\begin{split}
&\int \frac{d^2\overline{k}_\perp}{16\pi^3}
 \psi_D
 \left(\hat{x}^\prime(\bar{x},\,\xi),\,\hat{\VEC{k}}^\prime_\perp(\bar{\VEC{k}}_\perp\,,\bar{x}\,,\xi)\right)
 \psi_p
 \left(\tilde{x}(\bar{x},\,\xi),\,\tilde{\VEC{k}}_\perp(\bar{\VEC{k}}_\perp\,,\bar{x}\,,\xi)\right)
  \= \\
& \=
\frac{N_p N_D}{16\pi^2}
\frac{\left( \bar{x} + \xi \right) \, \left( \bar{x}^2 - \xi^2 \right) \, \left( 1 - \bar{x} \right)}{\left(1+\xi\right)}
\frac{1}{a_{D}^2 \left(1-\xi\right)^2\left(\bar{x}+\xi\right) + a_p^2 \left(1+\xi\right)^2\left(\bar{x}-\xi\right)} \\
& \times \exp\left[ - a_{D}^2 M^2 \frac{\left(\bar{x} - \xi
- x_0\left(1-\xi\right)\right)^2} {\left(\bar{x}-\xi\right)
\left(1-\bar{x}\right)}\right]
 \exp\left[- \bm{\Delta}_{\perp}^2 \frac{a_{D}^2 a_p^2 \left(1-\bar{x}\right)}
 {a_{D}^2 \left(1-\xi\right)^2
\left(\bar{x}+\xi\right) + a_p^2
\left(1+\xi\right)^2\left(\bar{x}-\xi\right)}\right] \,.
\end{split}
\label{Eq_GPD}
\ee
\begin{figure}[h]
\subfigure{
\includegraphics[width=.49\textwidth, clip=true]
{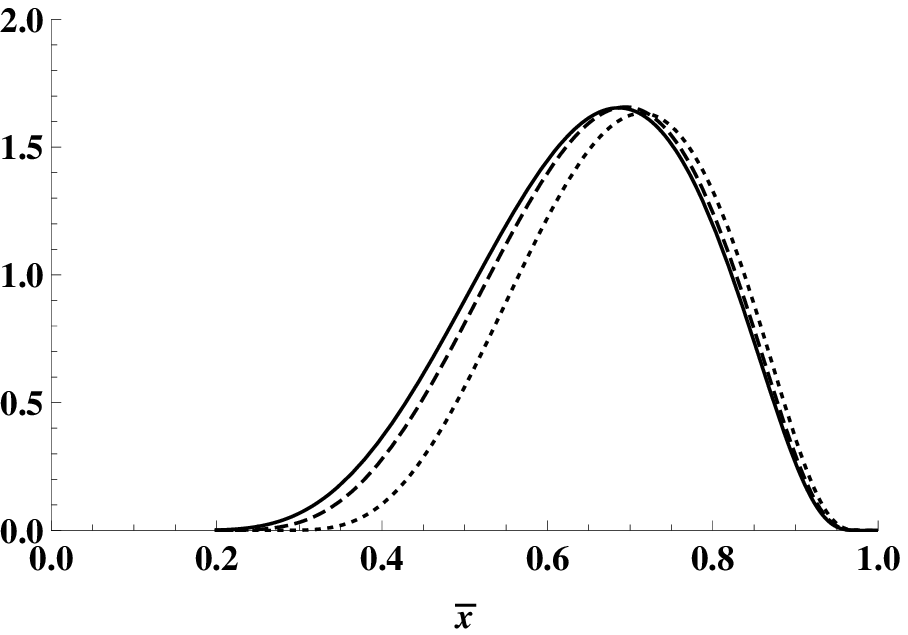}
}
\hfill
\subfigure{
\includegraphics[width=.49\textwidth, clip=true]
{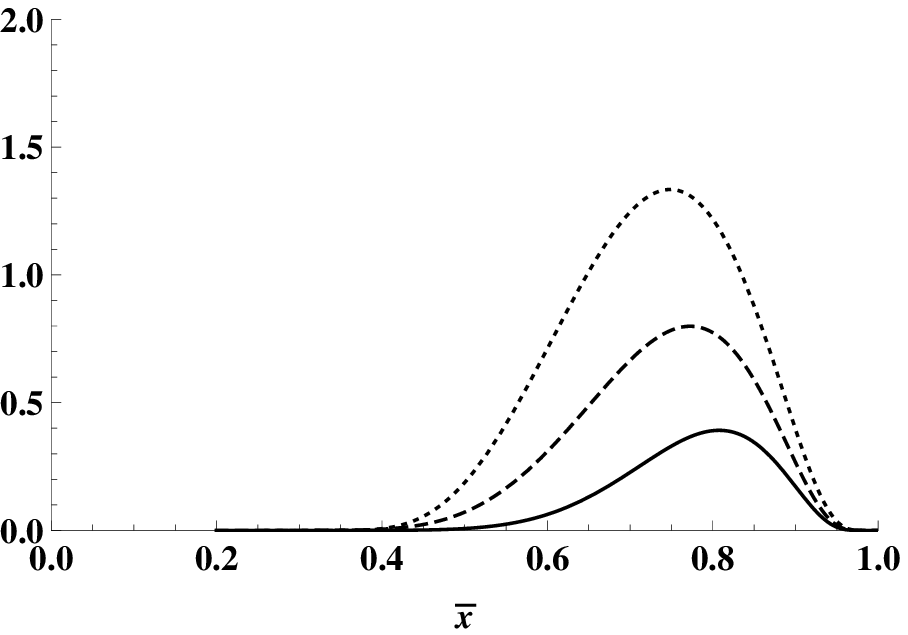}
}
\caption{
The wave function overlap of Eq.~\req{Eq_GPD} versus $\bar{x}$
at the CMS scattering angle $\theta = 0$ (upper figure)
and $\theta = \pi/2$ (lower figure).
We show it for Mandelstam $s = 30,\,20$ and $15\,\gev^2$
(solid, dashed and dotted curves).
}
\label{fig:H_vs_xbar}
\end{figure}
In Fig.~\ref{fig:H_vs_xbar} we show the wave function overlap
of Eq.~\req{Eq_GPD} versus the momentum fraction $\bar{x}$
with the parameters chosen as stated above.
First we observe that it is centered at $\bar{x} \approx x_0$
for vanishing CMS scattering angle.
Next, let us compare the upper and the lower panel. We see that the
magnitude of the wave function overlap is strongly decreasing with
increasing CMS scattering angle $\theta$.
The wave function overlap is also more pronounced in magnitude and shape in forward direction.
Furthermore, when comparing the overlap for different values of Mandelstam $s$,
we observe that in the more important forward scattering hemisphere the overlap is
increasing in magnitude with increasing CMS energy $s$, whereas at large scattering
angles this behavior is reversed.

\section{Cross Sections}
\label{sec:Cross-Sections}
%

The differential cross section for $\ourprocess$ reads
\be
 \frac{d\sigma_{\ourprocess}}{d\Omega}
 \=
\frac{1}{4\pi} s \Lambda_M \Lambda_m \frac{d\sigma_{\ourprocess}}{dt}
  \=
\frac{1}{64\pi^2} \frac{1}{s} \frac{\Lambda_M}{\Lambda_m} \sigma_0 \,,
\label{eq:diff-CS}
\ee
where we have introduced
\be
 \sigma_0 \,:=\, \frac14 \sum_{\mu\nu} \, \mid M_{\mu\nu} \mid^2 \,.
\label{eq:sigma0}
\ee

\begin{figure}[h!]
\subfigure{
\includegraphics[width=.49\textwidth, clip=true]
{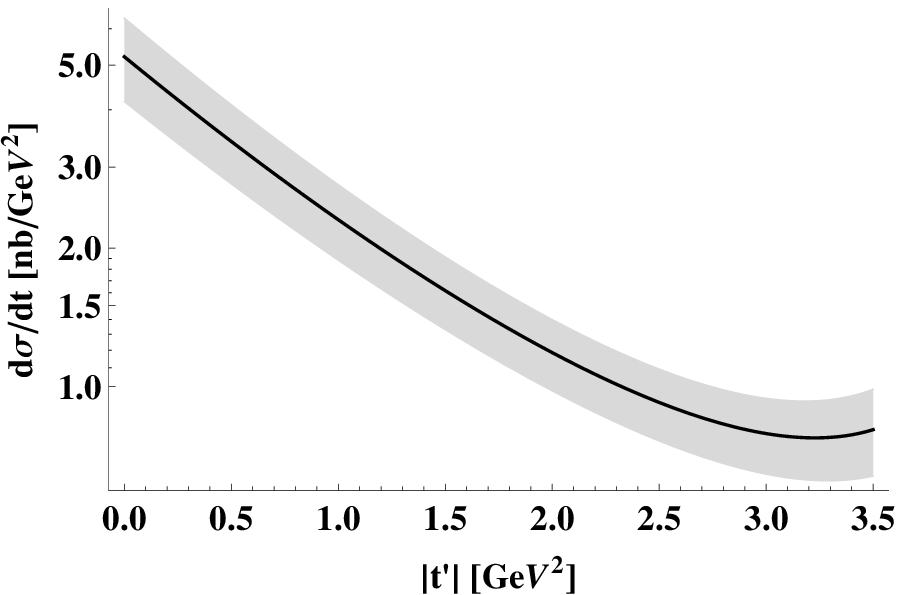}
}
\hfill
\subfigure{
\includegraphics[width=.49\textwidth, clip=true]
{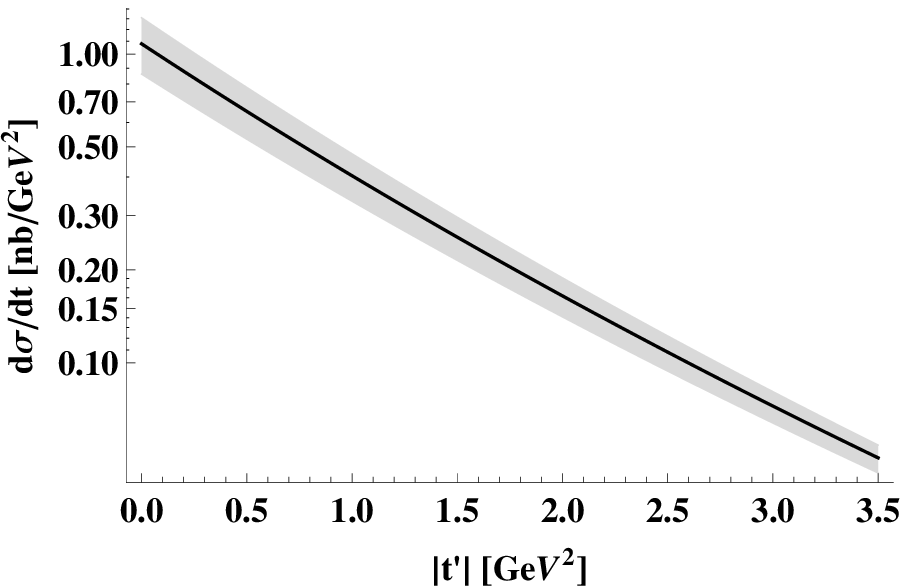}
}
\begin{center}
\subfigure{
\includegraphics[width=.49\textwidth, clip=true]
{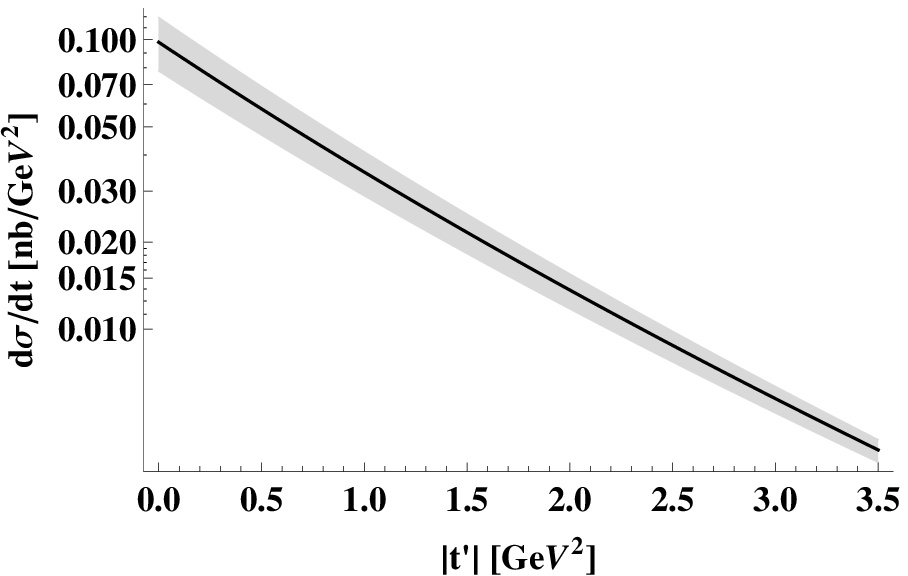}
}
\end{center}
\caption{
The differential cross section $d\sigma_{\ourprocess} / dt$
versus $\mid t^\prime \mid$. On the upper, middle and lower panel
we show it for Mandelstam $s = 15,\,20$ and $30\,\gev^2$, respectively.
}
\label{fig:dsdt}
\end{figure}
In Fig.~\ref{fig:dsdt} the differential cross section
$d\sigma_{\ourprocess}/dt$ is plotted versus $\mid t^\prime \mid$,
again for Mandelstam $s=15,\,20$ and $30\,\gev^2$.
The decrease of the cross section with increasing $\mid t^\prime \mid$
can mainly be attributed to the wave function overlap which
gives rise to a generalized form factor
[cf. Eq.~\req{Mmunu-overlap-final-peaking-approx}].
This form factor enters the differential cross section
to the fourth power.
The forward direction is dominated by those amplitudes in which
the helicities of the proton and antiproton (and also of the
$c$ and $\bar{c}$ quark) are equal.
They go with $\cos\theta$.
With increasing scattering angle they compete with those in which
proton and antiproton (and also $c$ and $\bar{c}$) have opposite helicities.
The latter go with $\sin\theta$ and dominate at $90^\circ$.
If one looks at the energy dependence one observes that
$M_{++}$ and $M_{--}$ are suppressed by a factor $2M/\sqrt{s}$ as
compared to $M_{+-}$ and $M_{-+}$
[cf. Eq.~\req{eq:subprocess-amplitudes}].
In the $p\bar{p} \,\to\, \Lambda_c^+ \overline{\Lambda_c}^-$ case of Ref.~\cite{Goritschnig:2009sq}
the factor $M/\sqrt{s}$ comes with those amplitudes which vanish in forward direction.
When comparing the different panels of Fig.~\ref{fig:dsdt} one sees that
the effect of the increase of the differential cross section with decreasing
scattering angle becomes more pronounced for higher CMS energies.

\begin{figure}[htb!]
\begin{center}
\includegraphics[clip=true]
{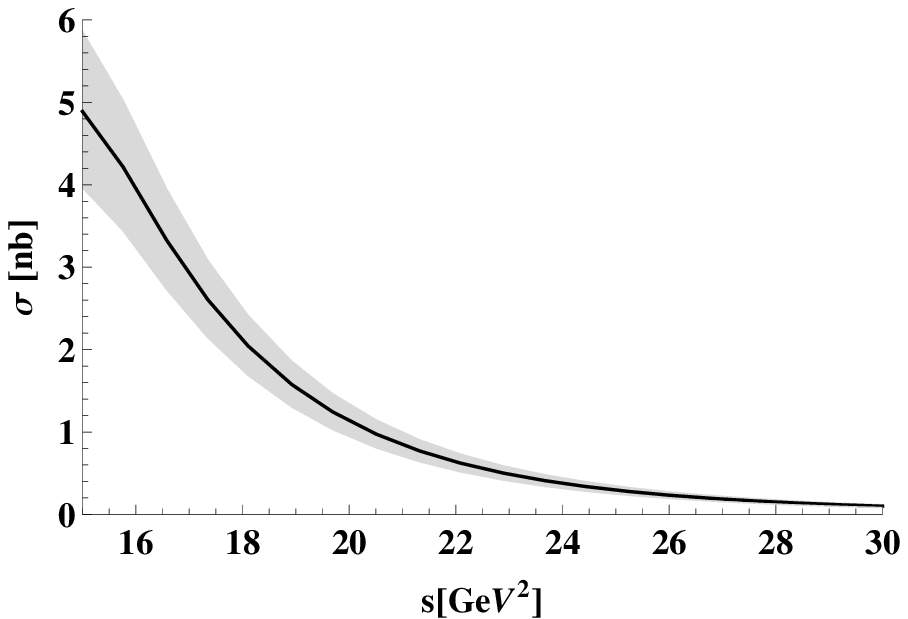}
\end{center}
\caption{
The integrated cross section $\sigma_{\ourprocess}$ versus Mandelstam $s$.
}
\label{fig:sigma}
\end{figure}

In Fig.~\ref{fig:sigma} we show the integrated cross section
$\sigma$ versus Mandelstam $s$.
It is of the order of $\text{nb}$, which is
of the same order of magnitude as the integrated cross section
for $p\bar{p} \,\to\, \Lambda_c^+ \overline{\Lambda_c}^-$ in
Ref.~\cite{Goritschnig:2009sq}.
This finding is in accordance with the diquark-model calculation
of Ref.~\cite{Kroll:1988cd}.
According to Ref.~\cite{Kroll:1988cd} larger cross sections
are to be expected for the  $p\bar{p} \,\to\, D^+ D^-$ reaction.
This, however, requires to extend our handbag approach by
including vector diquarks and will be the topic of
future investigations.
Our estimated cross section is about one order of magnitude
smaller than the predictions given in Refs.~\cite{Khodjamirian:2011sp, Haidenbauer:2010nx},
where hadronic interaction models have been used.
Whereas the authors of Ref.~\cite{Khodjamirian:2011sp}
determine their couplings of the initial proton
to the intermediate and final charmed hadrons by means of QCD sum rules,
the authors of Ref.~\cite{Haidenbauer:2010nx}
rather use $\text{SU}(4)$ flavor symmetry.
Though their predictions for the integrated $\ourprocess$ cross section
are comparable, they differ substantially in the
$p\bar{p} \,\to\, D^+ D^-$ cross section which, in
Ref.~\cite{Khodjamirian:2011sp}, is even smaller than our
$\ourprocess$ cross section.
Such big discrepancies reveal the high necessity of
experimental data which allow to decipher between
different dynamical models.
Such experiments could also help to pin down the charm-quark
content of the proton sea.
A considerably higher cross section within our approach could
only be explained if the charm-quark content of the proton sea
was not negligible.

\section{Summary}
\label{sec:Summary}
%
We have described the exclusive process $p\overline{p} \rightarrow \overline{D^0} D^0$ by means of a double handbag mechanism. This means that the process was assumed to factorize into a hard subprocess on the constituent level, which can be treated by means of perturbative QCD, and into soft hadronic matrix elements describing the nonperturbative $p\rightarrow \overline{D^0}$ and $\overline{p}\rightarrow D^0$ transitions. The intrinsic hard scale, justifying this approach, is given by the mass of the $c$ quark.

In order to produce the $\overline{D^0} D^0$ pair via $p\overline{p}$ annihilation a $u\overline{u}$ and a $d\overline{d}$ pair has to be annihilated on the constituent level and a $c\bar{c}$ pair must be created. We have adopted the simplifying assumption that the (dominant) valence Fock component of the proton consists of a scalar $S[ud]$ diquark and a $u$ quark such that the flavor changing hard process on the constituent level then becomes a simple $S[ud] \overline{S[ud]}\rightarrow c\overline{c}$ annihilation via the exchange of a highly virtual gluon. When calculating this annihilation, the composite nature of the $S[ud]$ diquark has been taken into account by a form factor at the diquark-gluon vertex. The form-factor parameter has been taken from the literature, where such kind of diquark model was already applied to other processes.

The soft part of the $p\rightarrow \overline{D^0}$ transition is encoded in a hadronic matrix element consisting of an $S[ud]$-diquark and a $c$-quark field operator, sandwiched between the incoming proton and the outgoing $\overline{D^0}$ state. We have given a parametrization of this matrix element in terms of (four) $p\rightarrow \overline{D^0}$ transition distribution amplitudes \cite{footnote1}. To model the $p\rightarrow \overline{D^0}$ transition matrix element we have employed an overlap representation in terms of light-cone wave functions of the proton and the $\overline{D^0}$. Such a representation makes sense for energies well above threshold and scattering angles in the forward hemisphere, where the momentum fractions of the active constituents have to be larger than the skewness. For the light-cone wave functions of the proton and the $\overline{D^0}$ we have taken simple oscillator-type models from the literature. The two parameters (normalization and oscillator parameter) in each case have been fixed such that the wave functions provide reasonable probabilities for the valence Fock state and reasonable values for the mean intrinsic transverse momentum (in case of the proton) and the $D^0$ decay constant. 

This overlap representation provided us with a model for the transition distribution amplitudes and allowed us to predict differential and integrated cross sections for the $p\overline{p} \rightarrow \overline{D^0} D^0$ process. For this simple wave function model only the transition distribution amplitude associated with the covariant $\gamma_5 u_{\mathrm{proton}}$ survived. The maximum size of the differential and integrated cross sections was found to be of the order of nb, i.e., about one order of magnitude smaller than corresponding cross sections calculated within hadronic interaction models. Experimental data are therefore highly needed to figure out the favorable approach. Higher cross sections can be expected for $p\overline{p} \rightarrow D^- D^+$ within our approach, but this would require to extend the concept of transition distribution amplitudes to vector diquarks.

\begin{acknowledgements}
We acknowledge helpful discussions with L.~Szymanowski.
This work was supported by the Austrian Science Fund FWF
under Grant No. J 3163-N16 and via the Austrian-French scientific
exchange program Amadeus.
\end{acknowledgements}


\begin{appendix}

\section{Kinematics}
\label{app-kinematics}

%
The four-momentum transfer $\Delta$ can be written as
[cf. (\ref{def-momenta-ji}), (\ref{sum-and-diff})]
\be
\Delta \=\ \Big[-2\xi\bar{p}^{\,+}\,,
  \frac{M^2(1+\xi)-m^2(1-\xi)+\xi\vdd/2}{2\bar{p}^{\,+}(1-\xi^2)}\,,
  \vd\Big]\,.
\label{Delta-explicit}
\ee
Note that $\Delta^+ = -\Delta^-$ since $p^\prime-p=q-q^\prime$.

In order to find expressions for the sine and the cosine of the CMS scattering angle $\theta$,
we write the absolute value of the three-momentum and
the momentum component into $z$ direction of the incoming proton as
\be
|\VEC{p}| \=\ \frac{\sqrt{s}}{2} \,\Lambda_m \, , \quad
p_3 \=\ \frac{\sqrt{s}}{2} \,\sqrt{\Lambda_m^2-\vdd/s}
\label{proton_three_momentum}
\ee
and that of the outgoing $\bar{D^0}$ as
\be
|\VEC{p}^\prime| \=\ \frac{\sqrt{s}}{2} \,\Lambda_M \, , \quad
|p_3^\prime| \=\ \frac{\sqrt{s}}{2} \,\sqrt{\Lambda_M^2-\vdd/s} \,,
\label{lambda_three_momentum}
\ee
respectively.
\noindent
Note that we have chosen the coordinate system in such a way that the $z$ component of
the incoming proton momentum is always positive.
But that of the outgoing  $\bar{D^0}$ can become negative at large scattering angles
due to the unequal-mass kinematics.
This change of sign occurs when $\bm{\Delta}_\perp^2$ reaches its maximal value
\be
\bm{\Delta}_{\perp\text{max}}^2 \=\ s \Lambda_M^2 \, ,
\ee
which follows directly from Eq.~(\ref{lambda_three_momentum}).
Then the CMS scattering angle can be written as
\be
\begin{split}
\theta &\=\ \arccos\left(\frac{p_3}{|\VEC{p}|}\right) + \arccos\left(\frac{p_3^\prime}{|\VEC{p}^\prime|}\right) \\
       &\=\ \arcsin\left(\frac{|\Delta_\perp|}{2|\VEC{p}|}\right) +
          \arcsin\left(\frac{|\Delta_\perp|}{2|\VEC{p}^\prime|}\right)\phantom{ + \pi} \,,
          \quad \text{for forward scattering}\\
       &\=\ \arcsin\left(\frac{|\Delta_\perp|}{2|\VEC{p}|}\right) -
          \arcsin\left(\frac{|\Delta_\perp|}{2|\VEC{p}^\prime|}\right) + \pi \,, 
          \quad \text{for backward scattering}.
\label{def_cms_theta}
\end{split}
\ee
\noindent
Using Eq.~(\ref{def_cms_theta}) the sine and cosine of the CMS scattering angle $\theta$ turn out to be
\be
\sin\theta \=\ \sqrt{\frac{\bm{\Delta}_\perp^2}{s}} \frac{1}{\Lambda_m\Lambda_M}
\left( \sqrt{\Lambda_m^2 - \frac{\bm{\Delta}_\perp^2}{s}} +
\text{sign}\left(p_3^\prime\right) \sqrt{\Lambda_m^2 - \frac{\bm{\Delta}_\perp^2}{s}} \right)
\label{sin_theta}
\ee
and
\be
\cos\theta \=\ \frac{\text{sign}\left(p_3^\prime\right)}{\Lambda_m\Lambda_M}
\left( \sqrt{\left( \Lambda_m^2 - \frac{\bm{\Delta}_\perp^2}{s} \right)
\left( \Lambda_M^2 - \frac{\bm{\Delta}_\perp^2}{s} \right)} -
\frac{1}{\text{sign}\left(p_3^\prime\right)} \frac{\bm{\Delta}_\perp^2}{s} \right),
\label{cos_theta}
\ee
respectively, where $\text{sign}\left(p_3^\prime\right)$ takes care of the kinematical situation of
forward or backward scattering.
Now we are able to express several kinematical variables in a compact form.
Starting from the definition (\ref{sum-and-diff}) of the average hadron momentum $\bar{p}$
and using Eqs.~(\ref{proton_three_momentum}), (\ref{lambda_three_momentum}),
(\ref{sin_theta}) and (\ref{cos_theta}) its plus component can be written as
\be
\begin{split}
\bar{p}^{\,+} &\=\ \frac12 \left( p^+ + p^{\prime +} \right)
           \=\ \frac{1}{2\sqrt{2}} \left( \left(p_0 + p_3\right) +
                                 \left(p^\prime_0 + p^\prime_3\right) \right) \\
          &\=\ \frac14\sqrt{\frac{s}{2}}
              \Big[2+ \sqrt{\Lambda_m^2-\vdd/s} +{\rm sign}(p^\prime_3)
              \sqrt{\Lambda_M^2-\vdd/s}\Big] \\
          &\=\ \frac14 \sqrt{\frac{s}{2}} \Big[2 + \sqrt{\Lambda_m^2 + \Lambda_M^2
              + 2\Lambda_m\Lambda_M \cos\theta } \Big]\,.
\end{split}
\ee
Note that in our symmetric CMS $\bar{q}^- = \bar{p}^+$.
For the skewness parameter $\xi$ we get
\be
\begin{split}
\xi &\=\ \frac{p^+ - p^{\prime +}}{p^+ + p^{\prime +}} \\
    &\=\ \frac{\sqrt{\Lambda_m^2-\vdd/s} -{\rm sign}(p^\prime_3)\sqrt{\Lambda_M^2-\vdd/s}}
        {2+ \sqrt{\Lambda_m^2-\vdd/s} +{\rm sign}(p^\prime_3)
        \sqrt{\Lambda_M^2-\vdd/s}} \\
    &\=\ \frac{\Lambda_m^2 - \Lambda_M^2}{\sqrt{\Lambda_m^2 +
        \Lambda_M^2 + 2 \Lambda_m\Lambda_M\cos\theta}} \,\,
        \frac{1}{2 + \sqrt{\Lambda_{m}^2 + \Lambda_{M}^2 + 2 \Lambda_{m}\Lambda_{M}\cos\theta}} \,.
\label{skew}
\end{split}
\ee
Note that, as a consequence of the unequal-mass kinematics, $\xi$ cannot become zero,
which is different from, e.g., Compton scattering
where $\xi$ would be equal to zero in such a symmetric frame.
For $p'_3\geq 0$, however, $\xi$ is fairly small in our case and tends to zero for $s\to\infty$. \\
\noindent
Now let us further investigate the Mandelstam variables and write them in a more compact form
with the help of Eqs.~(\ref{proton_three_momentum}), (\ref{lambda_three_momentum}),
(\ref{sin_theta}) and (\ref{cos_theta}).
Mandelstam $t$ can be written as
\be
\begin{split}
t &\ = \ -\frac{\vdd}{1-\xi^2} - \frac{2\xi}{1-\xi^2}\Big[(1+\xi) M^2
-(1-\xi) m^2\Big] \\
 &\ = \ -\frac{\vdd}{2}
         -\frac{s}{4}\Big[\Lambda_m^2 + \Lambda_M^2 - 2\text{sign}\left(p^\prime_3\right)
     \sqrt{\Lambda_m^2-\vdd/s}\sqrt{\Lambda_M^2 - \vdd/s}\Big] \\
 &\ = \ -\frac{s}{4} \Big[\Lambda_m^2 + \Lambda_M^2 - 2 \Lambda_m\Lambda_M\cos\theta \Big]\,.
\label{def:t}
\end{split}
\ee
It cannot become zero for forward scattering but acquires the value
\be
t_{0} \ := \ t(\vdd=0,p^\prime_3\geq 0) \ = \ -\frac{s}{4} \,(\Lambda_m - \Lambda_M)^2\,,
\label{min-t}
\ee
and for backward scattering
\be
t_{1} \ := \ t(\vdd=0,p^\prime_3\leq 0) \ = \ -\frac{s}{4} \,(\Lambda_m + \Lambda_M)^2\,.
\label{backwd-t}
\ee
It is furthermore convenient to introduce a \lq\lq reduced\rq\rq\ Mandelstam variable $t'$
that vanishes for forward scattering,
\be
\begin{split}
t^\prime \ := \ & t-t_0 \\
         \ = \ & -\frac{\vdd}{2}-\frac{s}{2}\Big[\Lambda_m\Lambda_M
               - {\rm sign}(p'_3)\sqrt{\Lambda_m^2-\vdd/s}\sqrt{\Lambda_M^2-\vdd/s}\Big] \\
         \ = \ & -\frac{s}{2} \Lambda_{m} \Lambda_{M} \left(1 - \cos\theta \right)\,.
\label{def:tprime}
\end{split}
\ee
\noindent
Also the transverse component of $\Delta$ can easily be written as a function of
the sine and the cosine of the scattering angle $\theta$ using Eqs.~(\ref{sin_theta}) and (\ref{cos_theta}),
\be
\vdd \=\ s \frac{\Lambda_m^2 \Lambda_M^2 \sin^2\theta}
            {\Lambda_m^2 + \Lambda_M^2 + 2 \Lambda_m\Lambda_M\cos\theta}\,,
\label{delta_perp_vs_theta}
\ee
or solving Eq.~\req{def:tprime} for $\vdd$ one finds
\be
\vdd \=\ -t^\prime\frac{s\Lambda_m\Lambda_M+t^\prime}{s/4(\Lambda_m+\Lambda_M)^2+t^\prime}.
\ee
If we define Mandelstam $u$ for forward scattering in an analogous way
\be
u_0 \ :=u \ (\vdd=0,p^\prime_3\geq 0) \ = \ -\frac{s}4\,(\Lambda_m+\Lambda_M)^2\,
\ee
and for backward scattering
\be
u_1 \ := \ u(\vdd=0,p^\prime_3\leq 0) \ = \ -\frac{s}4\,(\Lambda_m-\Lambda_M)^2\,,
\ee
the sine and the cosine of half the CMS scattering angle $\theta$ can be written compactly as
\be
\sin^2\,\Big(\frac{\theta}{2}\Big) \ = \ \frac{1-\cos{\theta}}{2} \=
  \frac{t_0-t}{s\Lambda_m\Lambda_M} \,,
\label{sin_theta_compact}
\ee
\be
\cos^2\,\Big(\frac{\theta}{2}\Big) \ = \ \frac{1+\cos{\theta}}{2} \=
  \frac{u_1-u}{s\Lambda_m\Lambda_M} \,,
\label{cos_theta_compact}
\ee
respectively.

\section{Light Cone Spinors}
\label{app-lc-spinors}

For our purposes we use the light cone spinors \cite{Soper:1972xc, Brodsky:1997de}.
They read
\be
u\left(p , \uparrow\right) =
\frac{1}{2^{1/4}}\frac{1}{\sqrt{p^+}}
\left(
\begin{array}{c}
p^+ + m/\sqrt{2} \\
p_\perp / \sqrt{2}  \\
p^+ - m/\sqrt{2} \\
p_\perp / \sqrt{2}
\end{array}
\right)
\quad , \quad
u\left(p , \downarrow\right) =
\frac{1}{2^{1/4}}\frac{1}{\sqrt{p^+}}
\left(\begin{array}{c}
-p_\perp^* / \sqrt{2} \\
p^+ + m/\sqrt{2} \\
 p_\perp^* / \sqrt{2}\\
-p^+ + m/\sqrt{2}
\end{array}\right)
 \,,
\label{eq:LC-Dirac-spinors-pos-z}
\ee
\be
v\left(p , \uparrow\right) =
\frac{-1}{2^{1/4}}\frac{1}{\sqrt{p^+}}
\left(\begin{array}{c}
-p_\perp^* / \sqrt{2} \\
p^+ - m/\sqrt{2} \\
 p_\perp^* / \sqrt{2}\\
-p^+ - m/\sqrt{2}
\end{array}\right)
\quad , \quad
v\left(p , \downarrow\right) =
\frac{-1}{2^{1/4}}\frac{1}{\sqrt{p^+}}
\left(\begin{array}{c}
p^+ - m/\sqrt{2} \\
p_\perp / \sqrt{2}  \\
p^+ + m/\sqrt{2} \\
p_\perp / \sqrt{2}
\end{array}\right)
 \,.
\label{eq:LC-Dirac-anti-spinors-pos-z}
\ee
for $p^3 > 0$ and
\be
u\left(p , \uparrow\right) =
\frac{1}{2^{1/4}}\frac{sign\left(p^1\right)}{\sqrt{p^-}}
\left(\begin{array}{c}
p_\perp^* / \sqrt{2} \\
p^- + m/\sqrt{2} \\
p_\perp^* / \sqrt{2} \\
p^- - m/\sqrt{2}
\end{array}\right)
\quad , \quad
u\left(p , \downarrow\right) =
\frac{1}{2^{1/4}}\frac{-sign\left(p^1\right)}{\sqrt{p^-}}
\left(\begin{array}{c}
p^- + m/\sqrt{2} \\
-p_\perp / \sqrt{2} \\
-p^- + m/\sqrt{2} \\
p_\perp / \sqrt{2}
\end{array}\right)
 \,,
\label{eq:LC-Dirac-spinors-neg-z}
\ee
\be
v\left(p , \uparrow\right) =
\frac{1}{2^{1/4}}\frac{sign\left(p^1\right)}{\sqrt{p^-}}
\left(\begin{array}{c}
p^- - m/\sqrt{2} \\
-p_\perp / \sqrt{2} \\
-p^- - m/\sqrt{2} \\
p_\perp / \sqrt{2}
\end{array}\right)
\quad , \quad
v\left(p , \downarrow\right) =
\frac{1}{2^{1/4}}\frac{-sign\left(p^1\right)}{\sqrt{p^-}}
\left(\begin{array}{c}
p_\perp^* / \sqrt{2} \\
p^- - m/\sqrt{2} \\
p_\perp^* / \sqrt{2} \\
p^- + m/\sqrt{2}
\end{array}\right)
\label{eq:LC-Dirac-anti-spinors-neg-z}
\ee
for $p^3 < 0$.
They satisfy the charge-conjugation relation
\be
v\left(p , \lambda\right) = \imath \gamma^2 u^*\left(p , \lambda\right)
\ee
and are normalized as
\be
\bar{u}_{\lambda^{\prime}}\left(p\right) u_{\lambda}\left(p\right) = 2m\delta_{\lambda^{\prime} \lambda}
\qquad\text{and}\qquad
\bar{v}_{\lambda^{\prime}}\left(p\right) v_{\lambda}\left(p\right) = -2m\delta_{\lambda^{\prime} \lambda}.
\ee
%

%
%
%
%

\section{TDAs}
\label{app-TDAs}

Following Ref.~\cite{PSS} the $p \to \overline{D^0}$ transition matrix element
can be decomposed at leading twist into the following covariant structures,
\be
\begin{split}
\Tilde{\mathcal{H}}^{\bar{c}S}_\mu
 \,:=\, & \bar{p}^{\,+}\,\int\frac{dz_1^-}{2\pi}\,e^{\imath \bar{x}_1 \bar{p}^+ z_1^-} \,
  \langle \overline{D^0}:\, p^\prime \mid
  \Psi^c_+\left(-z_1^-/2\right) \Phi^{S[ud]}\left(+z_1^-/2\right)
  \mid p:\, p,\,\mu \rangle \\
 \,=\, & \gamma_5 \, u(p,\,\mu) \, V_1(\bar{x}_1,\,\xi,\,t)
  + \,  \frac{\Delta\sla}{M+m} \gamma_5 \, u(p,\,\mu) \, V_2(\bar{x}_1,\,\xi,\,t)  \\
& + \, u(p,\,\mu) \, \Tilde{V}_1(\bar{x}_1,\,\xi,\,t)
  + \, \frac{\Delta\sla}{M+m} \, u(p,\,\mu) \, \Tilde{V}_2(\bar{x}_1,\,\xi,\,t) \,,
\label{eq:TDA-parameterization}
\end{split}
\ee
where we have introduced the $p \to \overline{D^0}$ TDAs
$V_1,\,V_2,\,\Tilde{V}_1$ and $\Tilde{V}_2$.
When evaluating the $\ourprocess$ amplitude the hadronic transition matrix element
(\ref{eq:TDA-parameterization}) appears within the spinor product
\be
\mathcal{H}^{\bar{c}S}_{\lambda_1^\prime\mu} \,=\,
\bar{v}(k_1^\prime,\,\lambda_1^\prime) \,\gamma^+\, \Tilde{\mathcal{H}}^{\bar{c}S}_\mu \,,
\label{eq:mathcalH}
\ee
cf. Eq.~(\ref{Mmunu}). Expressed in terms of TDAs we thus have
\be
\begin{split}
\mathcal{H}^{\bar{c}S}_{\lambda_1^\prime\mu}
\,=\, & \bar{v}(k_1^\prime,\,\lambda_1^\prime) \,\gamma^+ \gamma_5 \, u(p,\,\mu) \, V_1(\bar{x}_1,\,\xi,\,t)
  + \, \bar{v}(k_1^\prime,\,\lambda_1^\prime) \,\gamma^+ \frac{\Delta\sla}{M+m} \gamma_5 \, u(p,\,\mu) \, V_2(\bar{x}_1,\,\xi,\,t)  \\
& + \,\bar{v}(k_1^\prime,\,\lambda_1^\prime) \,\gamma^+ u(p,\,\mu) \, \Tilde{V}_1(\bar{x}_1,\,\xi,\,t)
  + \,\bar{v}(k_1^\prime,\,\lambda_1^\prime) \,\gamma^+ \frac{\Delta\sla}{M+m} \, u(p,\,\mu) \, \Tilde{V}_2(\bar{x}_1,\,\xi,\,t) \\
 \,=\, & \sqrt{\frac{\bar{x}_1 - \xi}{1 - \xi}} \,
  \left[ \bar{v}(p^\prime,\,\lambda_1^\prime) \,\gamma^+ \gamma_5 \, u(p,\,\mu) \, V_1(\bar{x}_1,\,\xi,\,t)
 + \, \bar{v}(p^\prime,\,\lambda_1^\prime) \,\gamma^+ \frac{\Delta\sla}{M+m} \gamma_5 \, u(p,\,\mu) \, V_2(\bar{x}_1,\,\xi,\,t) \right. \\
 & \left. + \,\bar{v}(p^\prime,\,\lambda_1^\prime) \,\gamma^+ u(p,\,\mu) \, \Tilde{V}_1(\bar{x}_1,\,\xi,\,t)
   + \,\bar{v}(p^\prime,\,\lambda_1^\prime) \,\gamma^+ \frac{\Delta\sla}{M+m} \, u(p,\,\mu) \, \Tilde{V}_2(\bar{x}_1,\,\xi,\,t) \right] \,,
\label{eq:mathcalHtilde-parameterization}
\end{split}
\ee
after making the replacement $k_1^\prime = x_1^\prime p^\prime$
in the $\bar{v}-$spinor.
Evaluating the various spinor products
which appear in Eq.~(\ref{eq:mathcalHtilde-parameterization})
by using the light cone spinors of Appendix~\ref{app-lc-spinors}
gives
\be
\mathcal{H}^{\bar{c}S}_{++} \=
\frac{4\bar{p}^{\,+}}{M+m}\sqrt{\frac{\bar{x}_1-\xi}{1+\xi}}\frac{\Delta_\perp}{2}
\left( V_2(\bar{x}_1,\,\xi,\,t)
+
\Tilde{V}_2(\bar{x}_1,\,\xi,\,t) \right) \,,
\label{eq:mathcalHupup}
\ee
\be
\mathcal{H}^{\bar{c}S}_{--} \=
\frac{4\bar{p}^{\,+}}{M+m}\sqrt{\frac{\bar{x}_1-\xi}{1+\xi}}\frac{\Delta_\perp}{2}
\left( V_2(\bar{x}_1,\,\xi,\,t)
-
\Tilde{V}_2(\bar{x}_1,\,\xi,\,t) \right)
\label{eq:mathcalHdndn}
\ee
and
\be
\begin{split}
\mathcal{H}^{\bar{c}S}_{+-} \= &
2\bar{p}^{\,+}\sqrt{\bar{x}_1-\xi}\sqrt{1+\xi}
\Big[
V_1(\bar{x}_1,\,\xi,\,t)
-
\Tilde{V}_1(\bar{x}_1,\,\xi,\,t) \\
& +
\frac{2\xi}{1+\xi} \frac{m}{M+m}
\left( V_2(\bar{x}_1,\,\xi,\,t)
+
\Tilde{V}_2(\bar{x}_1,\,\xi,\,t) \right)
\Big] \,,
\label{eq:mathcalHupdn}
\end{split}
\ee
\be
\begin{split}
\mathcal{H}^{\bar{c}S}_{-+} \= &
-2\bar{p}^{\,+}\sqrt{\bar{x}_1-\xi}\sqrt{1+\xi}
\Big[
V_1(\bar{x}_1,\,\xi,\,t)
+
\Tilde{V}_1(\bar{x}_1,\,\xi,\,t)  \\
&  +
\frac{2\xi}{1+\xi} \frac{m}{M+m}
\left( V_2(\bar{x}_1,\,\xi,\,t)
-
\Tilde{V}_2(\bar{x}_1,\,\xi,\,t) \right)
\Big] \,.
\label{eq:mathcalHdnup}
\end{split}
\ee
The TDAs $V_1,\,\Tilde{V}_1,\,V_2$ and $\Tilde{V}_2$ can now be
expressed as linear combinations of $\mathcal{H}^{\bar{c}S}_{++},\,
\mathcal{H}^{\bar{c}S}_{--},\,\mathcal{H}^{\bar{c}S}_{+-}$ and
$\mathcal{H}^{\bar{c}S}_{-+}$.
For our overlap representation of the hadronic
transition matrix elements we have
$\mathcal{H}^{\bar{c}S}_{++} = \mathcal{H}^{\bar{c}S}_{--} = 0$
[cf. Eq.~\req{eq:vbar-gammaplus-v-etc}].
This means that $V_2 = \Tilde{V}_2 = 0$ and
\be
V_1 \=
\frac{1}{4\bar{p}^{\,+}\sqrt{\bar{x}_1-\xi}\sqrt{1+\xi}}
\left(
\mathcal{H}^{\bar{c}S}_{+-} - \mathcal{H}^{\bar{c}S}_{-+} \,,
\right) \,,
\ee
\be
\Tilde{V}_1 \=
- \frac{1}{4\bar{p}^{\,+}\sqrt{\bar{x}_1-\xi}\sqrt{1+\xi}}
\left(
\mathcal{H}^{\bar{c}S}_{+-} + \mathcal{H}^{\bar{c}S}_{-+}
\right) \,.
\ee
We further have
$\mathcal{H}^{\bar{c}S}_{+-} = -\mathcal{H}^{\bar{c}S}_{-+}$
[cf. Eq.~\req{hadronic-transition-overlap}],
so that we finally get
\be
V_1 \=
\frac{1}{2\bar{p}^{\,+}\sqrt{\bar{x}_1-\xi}\sqrt{1+\xi}}
\mathcal{H}^{\bar{c}S}_{+-}
\quad\text{and}\quad
\Tilde{V}_1 \= 0 \,.
\ee

\end{appendix}


\end{document}